\documentclass[usenatbib]{mn2e}

\usepackage{aas_macros}
\usepackage{amsmath}
\usepackage{amssymb}
\usepackage[colorlinks=true,breaklinks=true,citecolor=blue]{hyperref} 
\usepackage{xspace}
\usepackage{color}
\usepackage{graphicx}
\usepackage{txfonts} 
\usepackage{mathtools} 
\usepackage{multirow}

\newcommand{\eg}{\mbox{e.\,g.,}\xspace} 

\newcommand{\ie}{\mbox{i.\,e.,}\xspace}

\renewcommand{\vec}[1]{\bmath{#1}} 

\newcommand{\HI}{\textsc{Hi}}

\definecolor{blue}{rgb}{0,0,1} 
\definecolor{yellow}{rgb}{1,0.84,0}
\definecolor{red}{rgb}{1,0,0}
\definecolor{mypink1}{rgb}{0.858, 0.188, 0.478}
\definecolor{darkgreen}{rgb}{0,0.5,0}
\definecolor{orange}{rgb}{1.0,0.4,0}

\title[3D Lyman-alpha forest and high density absorbers]{Correlations in the three-dimensional Lyman-alpha forest contaminated by high column density absorbers}
\author[K.~K.~Rogers et al.]
{Keir K.~Rogers,$^{1,2}$\thanks{E-mail: keir.rogers@fysik.su.se} Simeon Bird,$^{3,5}$ Hiranya V.~Peiris,$^{2,1}$ Andrew Pontzen,$^{2}$
\newauthor Andreu Font-Ribera$^{2}$ and Boris Leistedt$^{4,5}$
\\
$^1$Oskar Klein Centre for Cosmoparticle Physics, Stockholm University, AlbaNova, Stockholm SE-106 91, Sweden\\
$^2$Department of Physics \& Astronomy, University College London, Gower Street, London WC1E 6BT, UK\\
$^3$Department of Physics \& Astronomy, Johns Hopkins University, 3400 N.~Charles Street, Baltimore, MD 21218, USA\\
$^4$Department of Physics, New York University, 726 Broadway, New York, NY 10003, USA\\
$^5$NASA Einstein Fellow}

\begin{document}
\maketitle
\begin{abstract}
Correlations measured in three dimensions (3D) in the Lyman-alpha forest are contaminated by the presence of the damping wings of high column density (HCD) absorbing systems of neutral hydrogen (\HI; having column densities \(N(\HI) > 1.6 \times 10^{17}\,\mathrm{atoms}\,\mathrm{cm}^{-2}\)), which extend significantly beyond the redshift-space location of the absorber. We measure this effect as a function of the column density of the HCD absorbers and redshift by measuring 3D flux power spectra in cosmological hydrodynamical simulations from the Illustris project. Survey pipelines exclude regions containing the largest damping wings. We find that, even after this procedure, there is a scale-dependent correction to the 3D Lyman-alpha forest flux power spectrum from residual contamination. We model this residual using a simple physical model of the HCD absorbers as linearly biased tracers of the matter density distribution, convolved with their Voigt profiles and integrated over the column density distribution function. We recommend the use of this model over existing models used in data analysis, which approximate the damping wings as top-hats and so miss shape information in the extended wings. The simple ``linear Voigt model'' is statistically consistent with our simulation results for a mock residual contamination up to small scales (\(|\vec{k}| < 1\,h\,\mathrm{Mpc}^{-1}\)). It does not account for the effect of the highest column density absorbers on the smallest scales (\eg \(|\vec{k}| > 0.4\,h\,\mathrm{Mpc}^{-1}\) for small damped Lyman-alpha absorbers; HCD absorbers with \(N(\HI) \sim 10^{21}\,\mathrm{atoms}\,\mathrm{cm}^{-2}\)). However, these systems are in any case preferentially removed from survey data. Our model is appropriate for an accurate analysis of the baryon acoustic oscillations (BAO) feature. It is additionally essential for reconstructing the full shape of the 3D flux power spectrum.
\end{abstract}
\begin{keywords}
large-scale structure of universe -- cosmology: theory -- quasars: absorption lines
\end{keywords}

\section{Introduction}
\label{sec:intro}
Absorption lines of the Lyman-alpha forest can be mapped in three dimensions (3D) (\ie line-of-sight direction along the lengths of quasar spectra and transverse direction in the angular positions of the spectra on the sky) to trace the fluctuations in the cosmological density field. Correlations in the Lyman-alpha forest are a powerful probe of high redshifts (\(z > 2\)), before dark energy came to dominate the evolution of the Universe. In particular, measurement of the 3D correlations on large scales (separations \(r \sim 100\,\mathrm{Mpc}\,h^{-1}\)) in the Lyman-alpha forest allows a measurement of the baryon acoustic oscillations (BAO) in the distribution of matter at \(z \sim 2.3\) \citep{2011JCAP...09..001S,2013A&A...552A..96B,2013JCAP...04..026S,2013JCAP...03..024K,2015A&A...574A..59D,2017arXiv170200176B}. 3D correlations between the Lyman-alpha forest and the distribution of quasars have also been measured, including the detection of BAO \citep{2013JCAP...05..018F,2014JCAP...05..027F,2017arXiv170802225D}. This has been achieved thanks to the large number of quasar spectra from the Baryon Oscillation Spectroscopic Survey \citep[BOSS;][]{2011AJ....142...72E,2013AJ....145...10D} (157,783 were suitable for analysis in Data Release 12; DR12) and the large sky area they cover (the footprint in DR12 covers approximately one quarter of the sky). Consequently, Lyman-alpha forest analyses are no longer restricted to measurements of the one-dimensional flux power spectrum (along the line-of-sight only), which probes smaller-scale clustering (\(k_{||} > 0.1\,h\,\mathrm{Mpc}^{-1}\)) and constrains cosmological models that suppress small-scale power, \eg those containing massive neutrinos or warm dark matter \citep[WDM;][]{2005PhRvD..71j3515S,2015JCAP...11..011P,2017arXiv170201764I,2017arXiv170203314Y,2017arXiv170304683I,2017arXiv170309126A}.

Current measurements of the 3D correlations in the Lyman-alpha forest reconstruct the correlation function, where the BAO feature is most distinguishable. However, ongoing analyses in the extended Baryon Oscillation Sky Survey \citep[eBOSS;][]{2016AJ....151...44D} and future surveys like the Dark Energy Spectroscopic Instrument \citep[DESI;][]{2016arXiv161100036D,2016arXiv161100037D} will also measure its Fourier-space counterpart, the 3D flux power spectrum \citep{FontRiberaP3D}. A measurement of the 3D Lyman-alpha forest power spectrum will probe the full shape on a wide range of scales (\(0.01\,h\,\mathrm{Mpc}^{-1} < k < 1\,h\,\mathrm{Mpc}^{-1}\)). On large scales (\(k < 0.1\,h\,\mathrm{Mpc}^{-1}\)), the 3D flux power spectrum can be used to determine the cosmological geometry through the Alcock-Paczy\'nski test \citep{1979Natur.281..358A,1999ApJ...511L...5H,1999ApJ...518...24M,2003ApJ...585...34M}. The 3D forest power spectrum on large scales can also be used to study fluctuations in the ultraviolet (UV) ionising background \citep{2014PhRvD..89h3010P,2014ApJ...792L..34P}.

On smaller scales (\(0.1\,h\,\mathrm{Mpc}^{-1} < k < 1\,h\,\mathrm{Mpc}^{-1}\)), the 3D flux power spectrum adds complementary information to that from the 1D flux power spectrum. \citet{FontRiberaP3D} show that the 3D flux power spectrum for the BOSS survey is more constraining than the 1D counterpart up to a maximum \(k = 1\,h\,\mathrm{Mpc}^{-1}\). Only for \(k > 1\,h\,\mathrm{Mpc}^{-1}\) does the 1D power spectrum contain essentially all information. For future surveys such as DESI, where there will be a higher density of lines of sight, one may anticipate 3D information to even higher \(k\), underscoring the importance of working with the 3D spectrum wherever possible. This will provide more power to constrain cosmological models with additional components (\eg massive neutrinos, WDM or fuzzy dark matter), or modifications to a simple power-law primordial power spectrum (\eg running of the primordial spectral index). In addition to providing greater statistical power, the 3D flux power spectrum is sensitive to different systematics than the 1D flux power spectrum (\eg in correlations with metal absorption lines). (See \eg \citealt{2014JCAP...05..023F} for forecasts of the constraining power of the 3D flux power spectrum with DESI.)

As with the 1D Lyman-alpha forest flux power spectrum \citep{2005MNRAS.360.1471M,2017arXiv170608532R}, 3D correlations in the Lyman-alpha forest are biased by the presence in quasar spectra of high column density (HCD) absorbers and their associated broadened absorption lines \citep{2011MNRAS.415.2257M,2011JCAP...09..001S,2012JCAP...07..028F}. HCD absorbers are defined as regions of neutral hydrogen (\HI) gas with a column density \(N(\HI)\) exceeding \(1.6 \times 10^{17}\,\mathrm{atoms}\,\mathrm{cm}^{-2}\), and are usually identified with the gas in or around galaxies. They form at the peaks of the underlying density distribution and so cluster more strongly than the Lyman-alpha forest \citep{2012JCAP...11..059F,2017arXiv170900889P}. The absorption lines of the highest column density systems are broadened, with large damping wings causing absorption in the spectrum away from the physical location of the absorber. These wings have a characteristic Voigt profile, a convolution of a Gaussian profile (caused by Doppler broadening) and a Lorentzian profile (caused by natural or collisional broadening). They are traditionally sub-classified as either damped Lyman-alpha absorbers (DLAs; \(N(\HI) > 2 \times 10^{20}\,\mathrm{atoms}\,\mathrm{cm}^{-2}\)) or Lyman-limit systems (LLS; \(1.6 \times 10^{17}\,\mathrm{atoms}\,\mathrm{cm}^{-2} < N(\HI) < 2 \times 10^{20}\,\mathrm{atoms}\,\mathrm{cm}^{-2}\)), according to the width of their damping wings \citep{1986ApJS...61..249W}. However, as noted in \eg \citet{2005MNRAS.360.1471M,2012JCAP...07..028F,2017arXiv170608532R}, systems with \(N(\HI)\) exceeding \(1 \times 10^{19}\,\mathrm{atoms}\,\mathrm{cm}^{-2}\) have significant wings, which we classify as sub-DLAs. In Lyman-alpha forest analyses, it is usual to attempt to ``clip'' out HCD absorbers by identifying their damping wings in spectra, masking the central absorption region and then correcting the wings (\eg see \citealt{2013AJ....145...69L} for details of the process for BOSS DR9 spectra). Nonetheless, there is always a residual contamination of HCD absorbers, since the smallest damping wings are hard to identify amongst instrumental noise and indeed the superposed Lyman-alpha forest itself. Estimates of the upper limit in column density for this residual contamination range from \(10^{20}\) to \(10^{21} \mathrm{atoms}\,\mathrm{cm}^{-2}\) \citep[\eg][]{2017arXiv170200176B}. It is therefore necessary to model the effect of this residual contamination to allow for robust cosmological inference from the Lyman-alpha forest (\citealt{2017arXiv170200176B,2017arXiv170802225D} were the first to model this component in a 3D correlation analysis).

There is a small literature on modelling the effect of (residual) HCD absorbers on correlations in the 3D Lyman-alpha forest. In Appendix B of \citet{2011MNRAS.415.2257M}, a linear model for the 3D flux power spectrum of HCD absorbers convolved with their (Voigt) absorption profiles is considered, allowing for their auto-correlation and cross-correlation with the Lyman-alpha forest. They show that this ``linear Voigt model'' predicts that the cross-correlation is the dominant component of the HCD absorbers' correction to the 3D Lyman-alpha forest power spectrum. \citet{2012JCAP...07..028F} measure the effect of HCD absorbers on the 3D Lyman-alpha forest correlation function using mock (quasar) spectra (details of their generation are given in \citealt{2012JCAP...01..001F}). They find the cross-correlation of HCD absorbers and the Lyman-alpha forest to indeed be the dominant systematic error on the Lyman-alpha forest auto-correlation. They additionally identify as significant terms the HCD absorber auto-correlation and a three-point correlation between two Lyman-alpha forest modes and an HCD absorber mode.

An approximate model for HCD absorbers is used in the measurement of the 3D Lyman-alpha forest correlation function with BOSS DR12 spectra \citep{2017arXiv170200176B} and the cross-correlation with the quasar distribution \citep{2017arXiv170802225D}. It is (the Fourier transform of) a biased linear power spectrum, with separate bias and redshift space distortion parameters for HCD absorbers, convolved with a top-hat filter in real space (\ie a sinc function in Fourier space) to approximate the profiles of HCD absorbers (hereafter, the ``BOSS model''). The large-scale bias of dark matter halos hosting DLAs can be constrained through the cross-correlation of DLAs in spectra with the Lyman-alpha forest using BOSS spectra \citep{2012JCAP...11..059F,2017arXiv170900889P}. This halo bias can then be related to the absorber flux transmission bias.

In this study, we measure the effect of HCD absorbers on correlations in the 3D Lyman-alpha forest using the 3D flux power spectrum in cosmological hydrodynamical simulations for the first time. We use simulation boxes from the Illustris project \citep{2014Natur.509..177V,2015A&C....13...12N}, which have been shown to reproduce the observed column density distribution function and spatial clustering of HCD absorbers at the 95\% confidence level \citep{2014Natur.509..177V,2014MNRAS.445.2313B}. We measure the full anisotropic effect as a function of column density and redshift. We then consider how well the linear Voigt model characterises our results and identify the regimes where this simple model breaks down. We also compare this model to the approximate BOSS model discussed above. Our results will improve the robustness of modelling HCD absorbers and hence improve cosmological inference for future Lyman-alpha forest surveys (\eg eBOSS/DESI).

We briefly explain the theory of the models that we consider for the Lyman-alpha forest and HCD absorbers in \S~\ref{sec:theory}. In \S~\ref{sec:method}, our methodology in measuring the 3D flux power spectrum from hydrodynamical simulations and our modelling procedure are explained. We present our main results in \S~\ref{sec:results}. These results are discussed and compared to previous work in \S~\ref{sec:discussion} and in \S~\ref{sec:concs}, we draw our conclusions.

\section{Theory}
\label{sec:theory}

\subsection{Lyman-alpha forest}
\label{sec:forest_theory}

\begin{figure}
\includegraphics[width=\columnwidth]{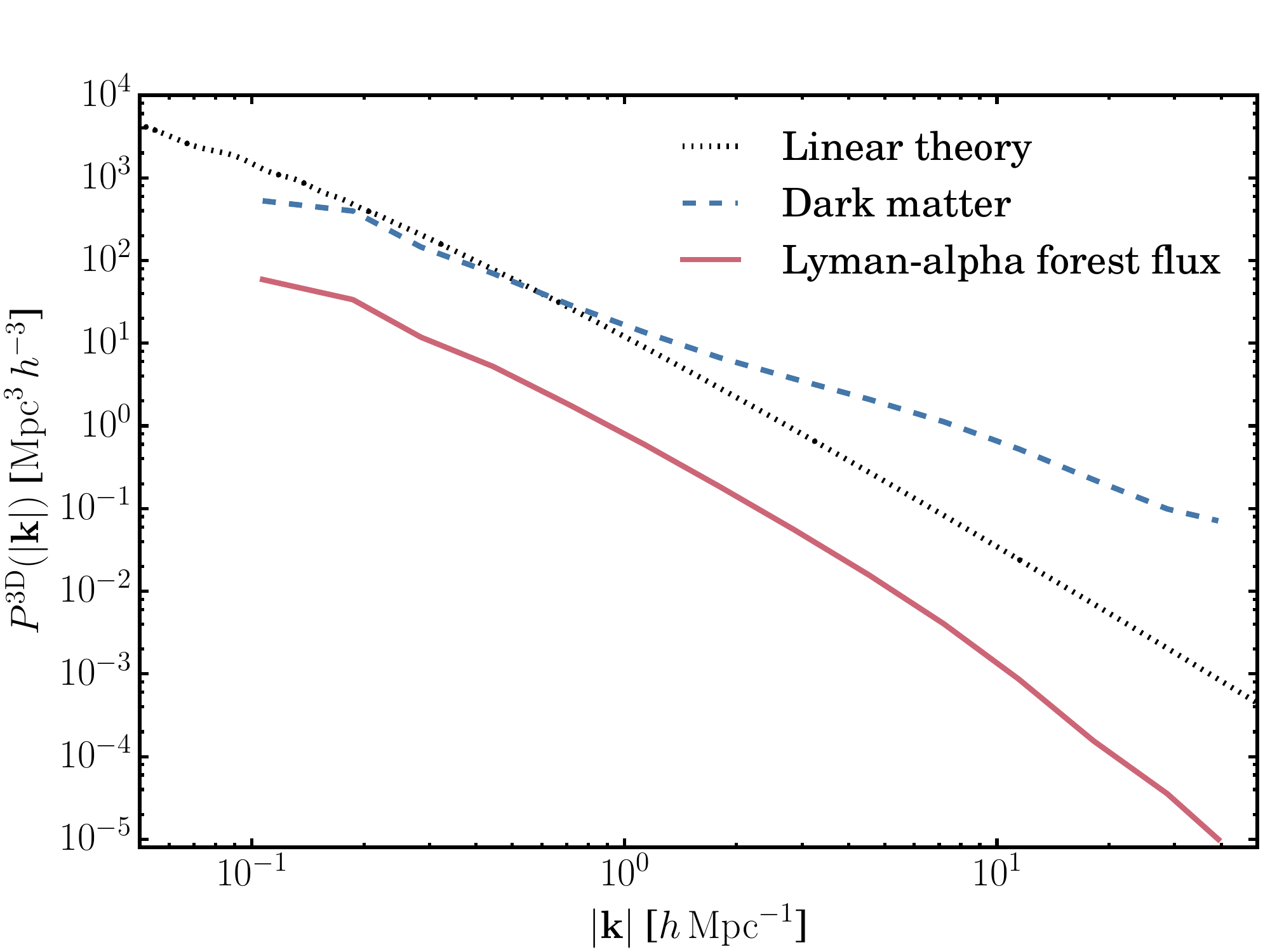}
\caption{A comparison of three-dimensional power spectra (averaged over all angles) as predicted by linear theory, measured from dark matter particles in a hydrodynamical simulation and measured from the transmission flux of the Lyman-alpha forest in redshift space from mock spectra generated from the same simulation. Although the Lyman-alpha forest is a biased tracer of linear theory, it remains linear to much smaller scales than other probes, including the power spectrum of dark matter, which is affected by non-linear gravitational evolution from scales larger than \(1\,h\,\mathrm{Mpc}^{-1}\). The simulation used is a \((75\,\mathrm{Mpc}\,h^{-1})^3\) box at redshift \(z = 2.44\) from the Illustris project \citep{2014Natur.509..177V}.}
\label{fig:linear_flux}
\end{figure}

Fluctuations in the transmitted flux of the Lyman-alpha forest are given as \(\delta_\mathrm{Forest}(\vec{x}) = \frac{\mathcal{F}_\mathrm{Forest}(\vec{x})}{\langle \mathcal{F}_\mathrm{Forest} \rangle} - 1\)\footnote{For the 3D comoving spatial coordinate \(\vec{x}\), the line-of-sight component \(x_{||}\) is transformed from the line-of-sight velocity space of spectra by the Hubble law.}, where the transmitted flux \(\mathcal{F}_\mathrm{Forest} = e^{- \tau_\mathrm{Forest}}\), \(\tau_\mathrm{Forest}\) is the optical depth and \(\langle \mathcal{F}_\mathrm{Forest} \rangle\) is the average flux over all spectral pixels. We then follow the standard treatment of the Lyman-alpha forest on large scales and model these fluctuations as a biased tracer of the underlying matter density fluctuation field with redshift-space distortions \citep[by analogy with other tracers of the matter distribution like galaxies or galaxy clusters;][]{1984ApJ...284L...9K,1987MNRAS.227....1K}. It therefore follows that the 3D Lyman-alpha forest flux power spectrum can be modelled as
\begin{equation}\label{eq:forest_linear}
P^\mathrm{3D}_\mathrm{Forest}(|\vec{k}|, \mu, z) = b^2_\mathrm{Forest} (1 + \beta_\mathrm{Forest} \mu^2)^2 P^\mathrm{3D}_\mathrm{Linear}(|\vec{k}|, z) D_\mathrm{NL}(|\vec{k}|, \mu),
\end{equation}
where \citep[following][]{2003ApJ...585...34M,2015JCAP...12..017A} we introduce a parametric function \(D_\mathrm{NL}(|\vec{k}|, \mu)\) to characterise deviations from linear theory due to non-linear effects; \(P^\mathrm{3D}_\mathrm{Linear}(|\vec{k}|, z)\) is the linear theory matter power spectrum; \(b_\mathrm{Forest}\) is the (linear) bias parameter of the Lyman-alpha forest; and \(\beta_\mathrm{Forest}\) is its redshift space distortion parameter. For the wavevector \(\vec{k}\) (conjugate to \(\vec{x}\)), we use a spherical coordinate system with its zenith direction along the line-of-sight such that power spectra are functions of \(|\vec{k}|\) and \(\mu\), which is the cosine of the angle between the wavevector and the line-of-sight. \(P^\mathrm{3D}_\mathrm{Forest}\) is a function of redshift \(z\) and in general, so are \(b_\mathrm{Forest}\) and \(\beta_\mathrm{Forest}\). Constraints on the redshift evolution of the Lyman-alpha forest power spectrum largely come at present from the 1D flux power spectrum (although there was some analysis of the redshift evolution of the 3D power spectrum in \citealt{2011JCAP...09..001S}). It is currently assumed \citep[\eg][]{2017arXiv170200176B} that \(b_\mathrm{Forest} \propto (1 + z)^\gamma\), where \(\gamma = 2.9\), and that \(\beta_\mathrm{Forest}\) does not depend on redshift such that \(b^2_\mathrm{Forest} P^\mathrm{3D}_\mathrm{Linear} \propto (1 + z)^{3.8}\), roughly matching the evolution observed in the 1D flux power spectrum \citep[\eg][]{2013A&A...559A..85P}. These assumptions are broadly supported by results from hydrodynamical simulations, although \(\beta_\mathrm{Forest}\) is found to decrease with increasing redshift \citep[\eg][]{2015JCAP...12..017A}.

The bias parameters of the Lyman-alpha forest differ from the biases of individual sources like galaxies or halos. For the latter, denser regions of matter contain a higher density of galaxies and halos and so their bias parameters are positive. For the forest, denser regions of matter will have less transmitted flux (due to increased absorption by {\HI} gas) and so \(b_\mathrm{Forest}\) is negative. Figure \ref{fig:linear_flux} compares the 3D Lyman-alpha forest flux power spectrum (averaged over all angles) as measured in redshift space from our simulation (see \S~\ref{sec:sims}) to the linear theory matter power spectrum. The flux power spectrum appears to be a scaled version of the theory power spectrum, remaining so to much smaller scales than, \eg the dark matter power spectrum, which is strongly affected by non-linear gravitational evolution for \(|\vec{k}| > 1\,h\,\mathrm{Mpc}^{-1}\). The deviation on small scales from linear theory is parameterised by the function \(D_\mathrm{NL}\), which is calibrated from hydrodynamical simulations. This function allows for the isotropic growth in power due to non-linear growth, isotropic suppression by pressure on very small scales and suppression by non-linear peculiar velocities and temperature towards the line-of-sight. For consistency with \citet{2017arXiv170200176B}, we use the fitting function of \citet{2003ApJ...585...34M} with the parameter values given in the first row of their Table 1. The parameters of this function have not been measured from data or previous simulations at the higher redshift that we consider (\(z = 3.49\)), and we are not able to do so with our simulations due to insufficient constraining power. Therefore for simplicity, we use the low-redshift parameter values when modelling the high-redshift setting. \citet{2015JCAP...12..017A}, in any case, found the shape of their simplest non-linear fitting function to evolve weakly in the redshifts they consider (\(2.2 \leq z \leq 3\)).

\subsection{High column density absorbers}
\label{sec:hcd_theory}

\begin{figure}
\includegraphics[width=\columnwidth]{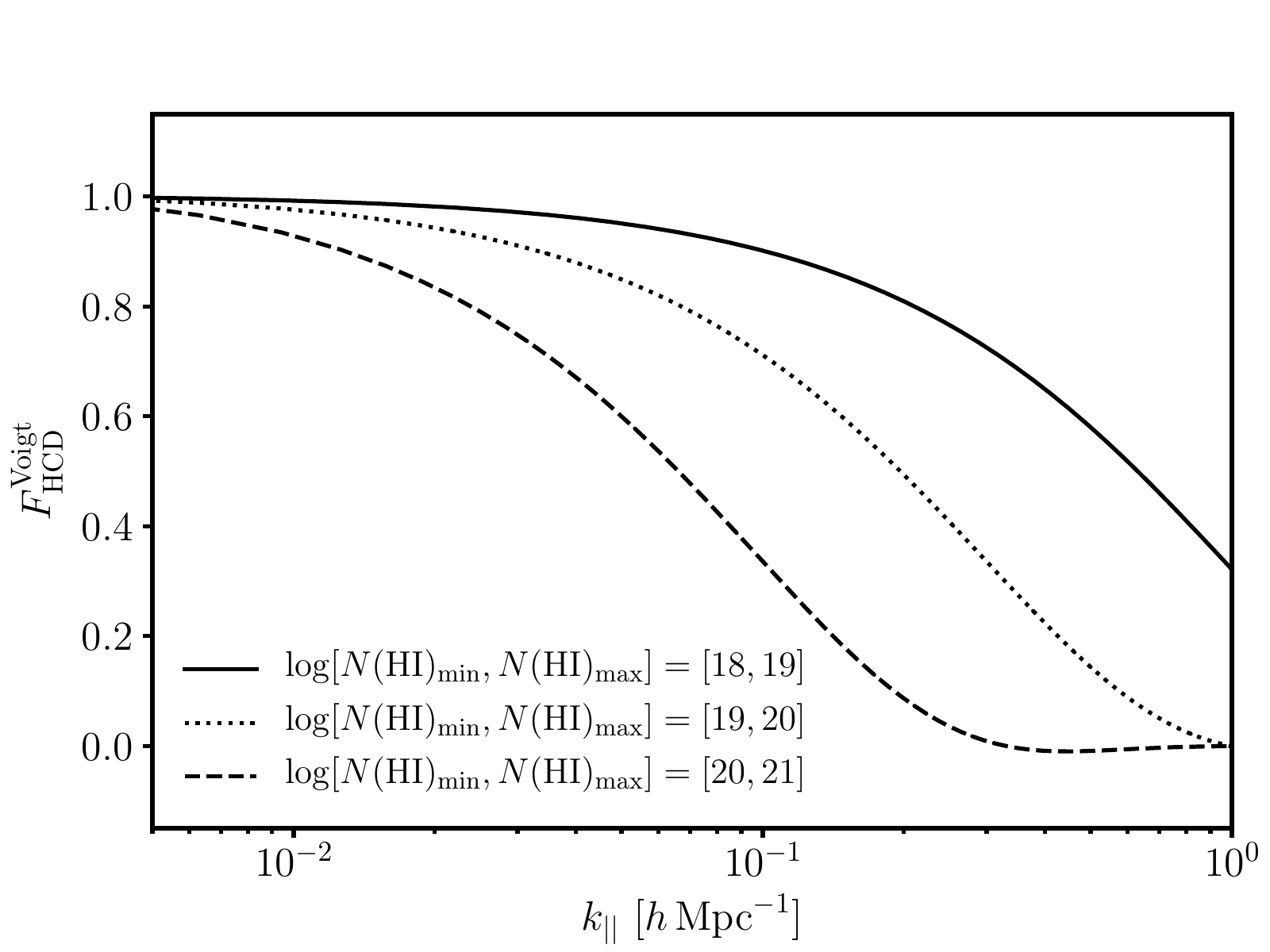}
\caption{\(F^\mathrm{Voigt}_\mathrm{HCD}(k_{||},z)\) (see Eq.~\eqref{eq:voigt_integral}) evaluated for the {\HI} CDDF (\(f(N(\HI),z)\)) as measured in our simulation box \citep[Illustris-1;][]{2014Natur.509..177V} at \(z = 2.44\). The units of \(N(\HI)\) are \(\mathrm{atoms}\,\mathrm{cm}^{-2}\).}
\label{fig:F_HCD}
\end{figure}

We follow \citet{2011MNRAS.415.2257M,2012JCAP...07..028F,2017arXiv170200176B} in modelling the 3D correlations of HCD absorbers on large scales. This is a linear model of HCD absorbers as point objects [\ie without damping wings; akin to Eq.~\eqref{eq:forest_linear}], convolved with the profiles of the wings. Therefore, the 3D flux power spectrum of a set of HCD absorbers with column densities in the interval \([N(\HI)_\mathrm{min}, N(\HI)_\mathrm{max}]\) is given as
\begin{equation}\label{eq:hcd_linear}
P^\mathrm{3D}_\mathrm{HCD}(|\vec{k}|, \mu, z) = b^2_\mathrm{HCD} (1 + \beta_\mathrm{HCD} \mu^2)^2 P^\mathrm{3D}_\mathrm{Linear}(|\vec{k}|, z) F^2_\mathrm{HCD}(k_{||},z),
\end{equation}
where \(b_\mathrm{HCD}\) and \(\beta_\mathrm{HCD}\) are the bias and redshift space distortion parameters of the absorption caused by HCD absorbers (these will in general depend on redshift \(z\)). \(F_\mathrm{HCD}\) is a function of the line-of-sight wavenumber \(k_{||} = |\vec{k}| \mu\), since it is caused by the absorption profiles of HCD absorbers which only manifest along the line-of-sight:
\begin{equation}\label{eq:voigt_integral}
F^\mathrm{Voigt}_\mathrm{HCD}(k_{||},z) = \int^{N(\HI)_\mathrm{max}}_{N(\HI)_\mathrm{min}} \mathrm{d}N(\HI) f(N(\HI),z) V(k_{||},N(\HI)).
\end{equation}
Here, \(V(k_{||},N(\HI))\) is the Fourier transform of the HCD absorbers' wing profiles as they manifest in the flux fluctuation field and \(f(N(\HI),z)\) is the column density distribution function (CDDF). The model we consider in this study uses the profile of HCD absorbers, which is a Voigt function in optical depth (see \eg Appendix A of \citealt{2017arXiv170608532R} for the full expression), the convolution of a Gaussian profile (caused by Doppler broadening) and a Lorentzian profile (caused by natural or collisional broadening). Fig.~\ref{fig:F_HCD} shows the shape of \(F^\mathrm{Voigt}_\mathrm{HCD}(k_{||},z)\) for some representative values of \([N(\HI)_\mathrm{min}, N(\HI)_\mathrm{max}]\). It shows that HCD absorbers of lower column density, which have narrower wings, have their effect on the power spectrum on smaller scales. We will also consider the approximation made by the BOSS Collaboration \citep{2017arXiv170200176B,2017arXiv170802225D} where the absorption profiles of HCD absorbers are modelled as top-hat filters [``BOSS model'']:
\begin{equation}\label{eq:boss_model}
F^\mathrm{BOSS}_\mathrm{HCD}(k_{||},z) = \frac{\sin(L_\mathrm{HCD} k_{||})}{L_\mathrm{HCD} k_{||}},
\end{equation}
where \(L_\mathrm{HCD}\) is a free parameter setting the effective width of these filters.

By combining Eqs.~\eqref{eq:forest_linear} and \eqref{eq:hcd_linear} (and additionally remembering the cross-correlation between the Lyman-alpha forest and HCD absorber fields), the 3D flux power spectrum for the Lyman-alpha forest contaminated by a set of HCD absorbers is given as:
\begin{equation}\label{eq:power_contam}
\begin{split}
P^\mathrm{3D}_\mathrm{Contaminated}(|\vec{k}|, \mu, z) = P&^\mathrm{3D}_\mathrm{Linear}(|\vec{k}|, z)\\
&[\tilde{b}^2_\mathrm{Forest} D_\mathrm{NL}(|\vec{k}|, \mu) + 2 \tilde{b}_\mathrm{Forest} \tilde{b}_\mathrm{HCD} + \tilde{b}^2_\mathrm{HCD}],
\end{split}
\end{equation}
where \(\tilde{b}_\mathrm{Forest} = b_\mathrm{Forest} (1 + \beta_\mathrm{Forest} \mu^2)\) and \(\tilde{b}_\mathrm{HCD} = b_\mathrm{HCD} (1 + \beta_\mathrm{HCD} \mu^2) F_\mathrm{HCD}(k_{||},z)\). If there was uncertainty in the CDDF of a given sample of spectra, it will be preferable to sub-divide the column density integrals evaluated in the calculation of \(F_\mathrm{HCD}\) in Eq.~\eqref{eq:voigt_integral} and allow for extra terms in Eq.~\eqref{eq:power_contam}, with bias parameters (\(\tilde{b}_{\mathrm{HCD},i}\)) for the \(N\) categories of HCD absorbers:
\begin{equation}\label{eq:power_extra}
\begin{split}
P^\mathrm{3D}_\mathrm{Contaminated}(|\vec{k}|, \mu, z) = P&^\mathrm{3D}_\mathrm{Linear}(|\vec{k}|, z) \left[\vphantom{\sum^N_{i = 1}}\tilde{b}^2_\mathrm{Forest} D_\mathrm{NL}(|\vec{k}|, \mu)\right.\\
&\left.+ \sum^N_{i = 1} \left(2 \tilde{b}_\mathrm{Forest} \tilde{b}_{\mathrm{HCD},i} + \sum^N_{j = 1} \tilde{b}_{\mathrm{HCD},i} \tilde{b}_{\mathrm{HCD},j}\right)\right].
\end{split}
\end{equation}
We mention also two possible additions that could be made to this model. First, the model in Eq.~\eqref{eq:power_contam} does not consider any non-linear evolution in the clustering of HCD absorbers\footnote{Indeed, the full ``BOSS model'' as used by \citet{2017arXiv170200176B,2017arXiv170802225D} multiplies the last two terms in Eq.~\eqref{eq:power_contam} by \(D_\mathrm{NL}\), the non-linear function calibrated by simulations of the Lyman-alpha forest only. We do not in the first instance include this correction to the linear Voigt model.}. Second, as noted in \citet{2012JCAP...07..028F}, in the two-point function of the total contaminated flux, there will arise three- and four-point functions of the Lyman-alpha forest and HCD absorber fluctuations. This is because the forest and HCD absorption terms are multiplied: \(\langle \mathcal{F}_\mathrm{Total} \rangle (1 + \delta_\mathrm{Total}) = \langle \mathcal{F}_\mathrm{Forest} \rangle (1 + \delta_\mathrm{Forest}) \langle \mathcal{F}_\mathrm{HCD} \rangle (1 + \delta_\mathrm{HCD})\), where \(\mathcal{F}_\mathrm{HCD}\) is the flux transmitted by HCD absorbers and \(\delta_\mathrm{HCD} = \mathcal{F}_\mathrm{HCD} / \langle \mathcal{F}_\mathrm{HCD} \rangle - 1\). It follows that in the total flux power spectrum (Eq.~\eqref{eq:power_contam}), there will be three- and four-point correlations involving \(\delta_\mathrm{Forest}\) and \(\delta_\mathrm{HCD}\). The model presented in Eq.~\eqref{eq:power_contam} only accounts for the leading two-point correlations; \citet{2012JCAP...07..028F}, however, found that the three-point term \(\langle \delta_\mathrm{Forest}(\vec{x_1}) \delta_\mathrm{HCD}(\vec{x_1}) \delta_\mathrm{Forest}(\vec{x_2}) \rangle\) is an important term on smaller scales (separations \(r < 40\,\mathrm{Mpc}\,h^{-1}\)). We discuss the possible impact of this additional term in \S~\ref{sec:discussion}.

\section{Method}
\label{sec:method}

We first outline the method we have used and then explain the steps in more detail in the following subsections (\S~\ref{sec:sims} to \ref{sec:mcmc_modelling}).
\begin{enumerate}
\renewcommand{\theenumi}{(\arabic{enumi})}
\item We use a cosmological hydrodynamical simulation from the Illustris project \citep{2014Natur.509..177V,2015A&C....13...12N} and generate mock spectra on a grid (562,500 in total, each at a velocity resolution of \(10\,\mathrm{km}\,\mathrm{s}^{-1}\) and with a typical length of \(\sim 8,000\,\mathrm{km}\,\mathrm{s}^{-1}\)). We calculate these at two redshift slices (\(z = [2.44, 3.49]\)). (See \S~\ref{sec:sims})
\item The mock spectra we generate contain absorption from the Lyman-alpha forest and HCD absorbers. For our analysis, it is useful to have a set of spectra containing only the Lyman-alpha forest (still forming a regular grid to allow the use of fast Fourier transforms; FFTs). To achieve this, we replace spectra contaminated by HCD absorbers by a nearby spectrum containing only the forest. Furthermore, we are able to construct boxes of spectra containing only the Lyman-alpha forest and a particular category of HCD absorber (\ie restricted to a particular column density interval) by replacing back the original spectra containing only that category of contamination. The details of this HCD ``dodging'' procedure are explained in \S~\ref{sec:hcd_dodging}.
\item For each box of spectra that we generate, we measure the three-dimensional (3D) flux power spectrum using an FFT. (See \S~\ref{sec:pow_spectrum}.)
\item Using these measurements of 3D flux power spectra, we fit the proposed model [Eq.~\eqref{eq:power_contam}] using a Markov chain Monte Carlo (MCMC) method. (See \S~\ref{sec:mcmc_modelling} and Appendix \ref{sec:mcmc_details}.)
\end{enumerate}

\subsection{Hydrodynamical simulations and mock spectra}
\label{sec:sims}

We use snapshots from the highest-resolution cosmological hydrodynamical simulation of the original Illustris project \citep[][Illustris-1\footnote{The simulation we use is publically available at \url{http://www.illustris-project.org/data}.}]{2014Natur.509..177V,2015A&C....13...12N}. The simulation adopts the following cosmological parameters: \(\Omega_\mathrm{m} = 0.2726\), \(\Omega_\Lambda = 0.7274\), \(\Omega_\mathrm{b} = 0.0456\), \(\sigma_8 = 0.809\), \(n_\mathrm{s} = 0.963\) and \(H_0 = 100\,h\,\mathrm{km}\,\mathrm{s}^{-1}\,\mathrm{Mpc}^{-1}\), where \(h = 0.704\) \citep{2014MNRAS.444.1518V}. The box has a comoving volume of \((106.5\,\mathrm{Mpc})^3\) and we consider snapshots at redshifts \(z = 2.44\) and \(3.49\). Illustris-1 has \(1820^3\) dark matter particles, each of mass \(6.3 \times 10^6\,\mathrm{M}_\odot\); the gas particle masses are each \(1.3 \times 10^6\,\mathrm{M}_\odot\). The simulations are in broad agreement with observations of the {\HI} CDDF \citep{2014Natur.509..177V}, the clustering of DLA halos \citep{2014MNRAS.445.2313B} and the kinematics of HCD absorbers \citep{2015MNRAS.447.1834B}. For a summary of the relevant physics in the simulation and comparisons to observations, see \citet{2017arXiv170608532R}. This broad agreement with relevant observations justifies the use of this simulation in building and testing models of HCD absorbers for use in future data analyses. This is further supported by the combination of large box size and high resolution. This allows the galaxy formation physics which is essential for correctly modelling HCD absorbers (which are generally associated with high-redshift galaxies) to be rendered in high resolution, while simultaneously allowing measurement of the large-scale behaviour. In particular, the large box size allows the generation of long mock spectra (see below), which can span the widths of large damping wings. This was not achievable with previous generations of simulations. A comparable simulation suite comes from the EAGLE project \citep{2015MNRAS.446..521S}, although their largest box is slightly smaller and has slightly lower mass resolution than Illustris-1.

For each snapshot, we generate mock spectra \citep[using the \texttt{fake\_spectra} code;][]{2017ascl.soft10012B} containing only the Lyman-alpha absorption line, on a square grid of \(750^2 = 562,500\) spectra, giving a spacing of 142 kpc between neighbouring spectra along each axis of the simulation box. Each spectrum extends the full length of the simulation box with periodic boundary conditions, giving a size in velocity space of \(7,501\) and \(8,420\,\mathrm{km}\,\mathrm{s}^{-1}\) respectively at \(z = 2.44\) and \(3.49\). We measure the optical depth \(\tau\) in velocity bins of size \(10\,\mathrm{km}\,\mathrm{s}^{-1}\) along the spectrum. We then calculate the transmitted flux \(\mathcal{F} = e^{-\tau}\). We convolve our spectra with a Gaussian kernel of \(\mathrm{FWHM} = 8\,\mathrm{km}\,\mathrm{s}^{-1}\), setting the simulated spectrographic resolution.

\subsection{Dodging high column density absorbers}
\label{sec:hcd_dodging}

We associate with each mock spectrum that is contaminated by HCD absorbers a nearby spectrum containing only Lyman-alpha forest absorption. Indeed, if the transverse distance necessary to ``dodge'' the contaminating HCD absorber is small \citep[as is expected considering the physical sizes of HCD absorbers,][]{2012MNRAS.424L...1K}, the large-scale cosmological modes (\ie the Lyman-alpha forest modes) in the replacement spectrum should be identical to the original spectrum and the difference will be the HCD absorber modes only. The ``dodging'' procedure iteratively proposes a nearby replacement spectrum until one is found with no HCD absorber contamination (\ie there are no column densities, integrated over \(100\,\mathrm{km}\,\mathrm{s}^{-1}\)\,\footnote{This is the same integration length as we used in our 1D flux power spectrum analysis in \citet{2017arXiv170608532R} and it amounts to ten neighbouring bins or a comoving length much larger than the most extensive HCD absorbers \citep{2012MNRAS.424L...1K}. In \citet{2017arXiv170608532R}, we also tested our sensitivity to the size of this integration length and found it made negligible difference to power spectrum estimates.}, exceeding \(1.6 \times 10^{17} \mathrm{atoms}\,\mathrm{cm}^{-2}\), the threshold for HCD absorbers). It searches for replacement spectra by successively generating spectra further away in a transverse direction from the original spectrum in steps of \(10\,\mathrm{kpc}\,h^{-1}\) until a suitable spectrum is found. In this way, we are able to generate a box of spectra containing only the Lyman-alpha forest.

We are also able to generate boxes of spectra containing Lyman-alpha forest and HCD absorbers of a certain category (\ie column densities in a certain interval) by replacing back original spectra containing this particular category. We categorise spectra according to the maximum column density (again integrated over \(100\,\mathrm{km}\,\mathrm{s}^{-1}\)) in each spectrum; there may be less dense HCD absorbers in each category but their effect will be sub-dominant since their damping wings are narrower.

Having generated these new boxes of spectra, we compute FFTs (\S~\ref{sec:pow_spectrum}), ignoring the transverse dodging distances and assuming that the dodged spectra lie on the original grid. Since the dodging distances are in general small (we find only \(\sim 1 \%\) to be \(> 500\,\mathrm{kpc}\,h^{-1}\); see Appendix \ref{sec:test_dodging}), the error associated with this approximation is restricted to small scales, \ie large \(|\vec{k}|\). We conduct an analysis of the error that arises from the irregular grid resulting from the dodging distances in Appendix \ref{sec:test_dodging}. Following these tests, we study only scales \(|\vec{k}| < |\vec{k}|_\mathrm{max}\), where \(|\vec{k}|_\mathrm{max} = 1\,h\,\mathrm{Mpc}^{-1}\); at these small values of \(|\vec{k}|\), the dodging error is negligible compared to the effect of HCD absorbers that we wish to measure (see Fig.~\ref{fig:fractional_hcd_effect}).

\subsection{Three-dimensional flux power spectrum}
\label{sec:pow_spectrum}

\begin{table}\centering
\caption{The neutral hydrogen ({\sc Hi}) column density limits \([N(\textsc{Hi})_\mathrm{min}, N(\textsc{Hi})_\mathrm{max}]\) that define the categories of absorbing systems used in this work. The columns on the right show the percentage of spectra (at each redshift \(z\) that is considered) in our \((106.5\,\mathrm{Mpc})^3\) simulation box \citep[][Illustris-1]{2014Natur.509..177V,2015A&C....13...12N} where the highest-density system belongs to a given category.}
\label{tab:hcd_absorbers}
\begin{tabular}{ccccc}
\hline
\multirow{2}{*}{Absorber category} & \(N(\textsc{Hi})_\mathrm{min}\) & \(N(\textsc{Hi})_\mathrm{max}\) & \multicolumn{2}{c}{\% of spectra in box at} \\
 & \multicolumn{2}{c}{[\(\mathrm{atoms}\,\mathrm{cm}^{-2}\)]} & \(z = 2.44\) & \(z = 3.49\) \\
\hline
Lyman-\(\alpha\) forest & 0 & \(1.6 \times 10^{17}\) & 69.6 & 45.7 \\
LLS & \(1.6 \times 10^{17}\) & \(1 \times 10^{19}\) & 14.9 & 27.0 \\
Sub-DLA & \(1 \times 10^{19}\) & \(2 \times 10^{20}\) & 8.1 & 14.3 \\
Small DLA & \(2 \times 10^{20}\) & \(1 \times 10^{21}\) & 4.1 & 7.8 \\
Large DLA & \(1 \times 10^{21}\) & \(\infty\) & 3.3 & 5.2 \\
\hline
\end{tabular}
\end{table}

We measure the 3D flux power spectrum at each redshift slice for our Lyman-alpha forest box of spectra (\(P^\mathrm{3D}_\mathrm{Forest}\)) and for our contaminated boxes of spectra for a number of HCD absorber categories (\(P^\mathrm{3D}_\mathrm{Contaminated}\)), the column density ranges of which we give in Table \ref{tab:hcd_absorbers}. We estimate the 3D flux power spectrum in bins of \(|\vec{k}|\) (15 bins) and \(\mu\) (4 bins), \(P^\mathrm{3D}_{\mathrm{Flux},i} = \frac{1}{N_i} \sum_n |\hat{\delta}_\mathrm{Flux}(\vec{k}_n)|^2\), where \(\vec{k}_n\) lie within a given \((|\vec{k}|,\mu)\) bin and \(N_i\) are the number of modes in each bin \(i\). \(\hat{\delta}_\mathrm{Flux}(\vec{k})\) is the Fourier transform of the flux fluctuation field \(\delta_\mathrm{Flux}(\vec{x}) = \frac{\mathcal{F}(\vec{x})}{\langle \mathcal{F} \rangle} - 1\). Here, for the mean flux \(\langle \mathcal{F} \rangle\) we always use the mean flux of the original box of spectra (with no dodging) so that our modelling assumption that the flux fluctuations can be sub-divided into different absorber categories \(\delta_\mathrm{Total} = \sum_i \delta_i\) holds true, where \(i\) indexes the different absorber categories. We use the convention of absorbing the \((2 \pi)^3\) into the conjugate variable, \ie we define the Fourier transform as \(\delta(k) = \int \delta(x) e^{-ikx} \mathrm{d}x\).

\subsection{Modelling and Markov chain Monte Carlo sampling}
\label{sec:mcmc_modelling}

We optimise the parameters of our model (the Lyman-alpha forest and HCD contamination biases and redshift-space distortion parameters) using MCMC sampling. We use MCMC sampling in order to estimate parameter uncertainties and to understand parameter degeneracies. Our data vectors consist of the flux power spectra of a contaminated set of spectra and an uncontaminated set (containing only Lyman-alpha forest). We model each data point as chi square-distributed (as is explained in Appendix \ref{sec:mcmc_details}) with an estimated variance of \(2 (P^\mathrm{3D}_{\mathrm{True},i})^2 / N_i\), where \(P^\mathrm{3D}_{\mathrm{True},i}\) is the true (ensemble) value of the 3D flux power spectrum in bin \(i\) and \(N_i\) is the number of modes per bin. These variances form the elements of our diagonal covariance matrices. The full details of the construction of our likelihood function and prior probability distributions are given in Appendix \ref{sec:mcmc_details}. The results are shown in \S~\ref{sec:modelling}.

\section{Results}
\label{sec:results}

\subsection{Measuring the effect of HCD absorbers}
\label{sec:measurements}

\begin{figure}
\includegraphics[width=\columnwidth]{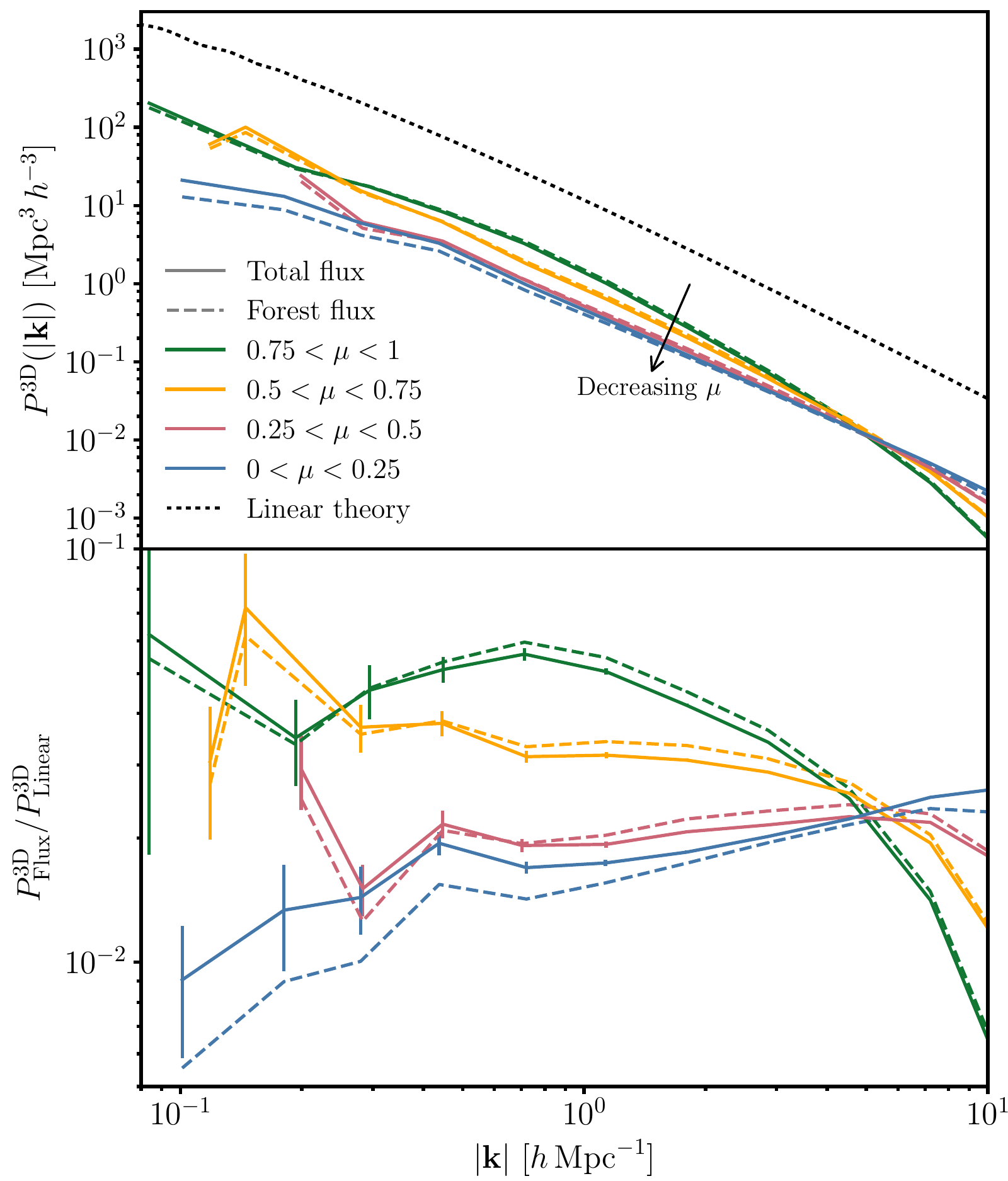}
\caption{\textit{Above}: the three-dimensional power spectra of the total flux from the Lyman-alpha forest and HCD absorbers (solid lines); and of the flux from the Lyman-alpha forest only (dashed lines); and the total linear theory matter power spectrum. For the flux power spectra, we show the anisotropic behaviour as a function of \(\mu\). \textit{Below}: the flux power spectra in ratio to the linear power spectrum. Flux measurements are made from a simulation box at redshift \(z = 2.44\).}
\label{fig:flux_power}
\end{figure}

Figure \ref{fig:flux_power} shows the measured 3D flux power spectra as a function of scale \(|\vec{k}|\) and the cosine of the angle away from the line-of-sight \(\mu\); \ie \(\mu = 1\) is along the line-of-sight and \(\mu = 0\) is transverse to the line-of-sight. Anisotropic behaviour arises due to linear redshift-space distortions on larger scales, enhancing power towards the line-of-sight. Non-linear effects on smaller scales suppress power along the line-of-sight due to non-linear peculiar velocities and thermal broadening of absorption lines. The non-linear effects are more manifest in the bottom panel, where ratios to the linear matter power spectrum are shown. On the largest scales, these ratios should tend towards constant values (\ie \(b^2 (1 + \beta \mu^2)^2\) for the linear models presented in \S~\ref{sec:theory}). However, we are only able to probe a small number of these large scale modes in our \(75\,\mathrm{Mpc}\,h^{-1}\) simulation box and so our measurement of large-scale bias has a large variance. The isotropic enhancement of power due to non-linear collapse of structure is broadly observable on larger scales (\(|\vec{k}| \sim 0.03\,h\,\mathrm{Mpc}^{-1}\)), although this trend is also obscured by the large variance on large scales. The anisotropic suppression of power towards the line-of-sight mentioned above is clearly observable for scales \(|\vec{k}| > 1\,h\,\mathrm{Mpc}^{-1}\), leading to the characteristic cross-over in the curves on small scales \citep{2003ApJ...585...34M,2015JCAP...12..017A}.

Figure \ref{fig:flux_power} also compares the 3D flux power spectra of contaminated and uncontaminated Lyman-alpha forest absorption (solid and dashed lines respectively; see \S~\ref{sec:hcd_dodging} for more details about how a box of spectra without HCD absorber contamination is constructed by the dodging technique). The contamination by HCD absorbers adds power in some regimes (especially in the transverse direction) and suppresses power in others (especially on smaller scales towards the line-of-sight). Following the tests of the error caused by the dodging procedure in forming the Lyman-alpha forest box of spectra (see Appendix \ref{sec:test_dodging} and \S~\ref{sec:hcd_dodging}), we cut our data-vectors at \(|\vec{k}|_\mathrm{max} = 1\) and we will throw away smaller scales in our following analysis. The forest flux power spectra shown in Fig.~\ref{fig:flux_power} (dashed lines) do not quantify the additional systematic and statistical error arising from the dodging (which is only significant for \(|\vec{k}| > 1\,h\,\mathrm{Mpc}^{-1}\)).

\begin{figure}
\includegraphics[width=\columnwidth]{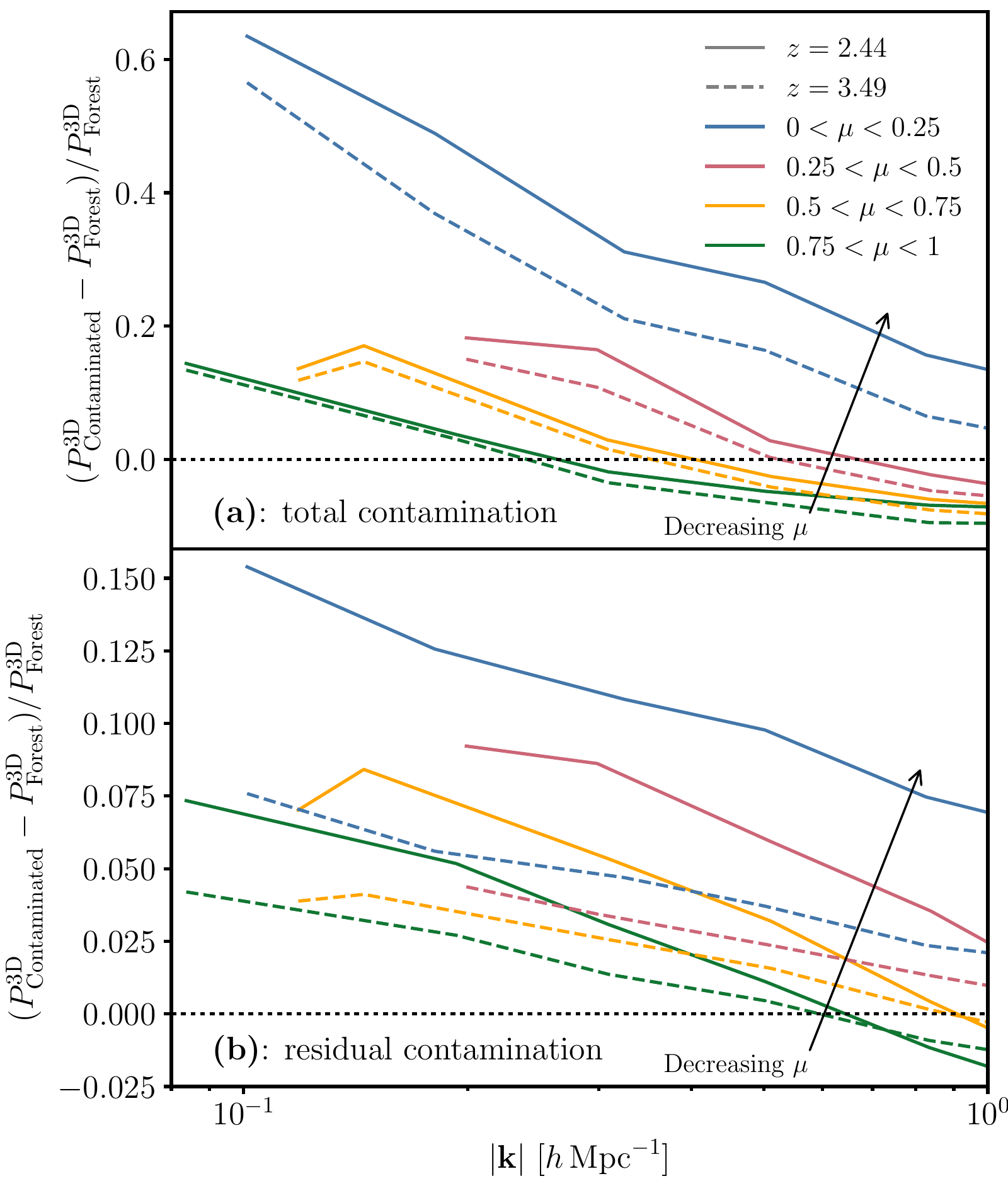}
\caption{The change in the flux power spectrum from contamination of quasar spectra by HCD absorbers, as a fraction of the Lyman-alpha forest power spectrum. \textbf{(a)} \textit{above}: the effect of the total contamination from all HCD absorbers in our simulation box; \textbf{(b)} \textit{below}: the effect of a mock residual contamination after the largest HCD absorbers have been ``clipped'' from quasar spectra (\ie only LLS and sub-DLAs remaining).}
\label{fig:fractional_hcd_effect}
\end{figure}

Figure \ref{fig:fractional_hcd_effect} shows the fractional effect of HCD absorber contamination on the 3D Lyman-alpha forest flux power spectrum. The fractional effect of the full ensemble of HCD absorbers [panel \textit{(a)}] can be as large as a 60\% correction to \(P^\mathrm{3D}(|\vec{k}|)\) at \(|\vec{k}| = 0.1\,h\,\mathrm{Mpc}^{-1}\) in the transverse direction. The fractional effect is smaller at higher redshift because the Lyman-alpha forest power spectrum (in the denominator) has a larger amplitude (since neutral hydrogen is more abundant and so there is stronger Lyman-alpha absorption). There is a larger fractional effect in the transverse direction driven also by the Lyman-alpha forest power spectrum, which has less power in this direction due to redshift-space distortions. The scale-dependence in Fig.~\ref{fig:fractional_hcd_effect} is partly driven by the non-linear effects in the Lyman-alpha forest power spectrum discussed above; in particular, the Lyman-alpha forest power spectrum is boosted on small scales due to non-linear growth and so the fractional effect decreases.

The bottom panel Fig.~\ref{fig:fractional_hcd_effect} \textit{(b)} shows the equivalent effects but for a mock residual contamination of HCD absorbers after the largest HCD absorbers have been ``clipped'' out (\ie only LLS and sub-DLAs remaining). The same trends are observed as above, but the overall amplitude is smaller since the largest damping wings have been removed. Nonetheless, the effect at \(|\vec{k}| = 0.1\,h\,\mathrm{Mpc}^{-1}\) in the transverse direction still constitutes a 15\% correction; it is therefore necessary to model this effect for robust cosmological inference from the Lyman-alpha forest (see \S~\ref{sec:modelling}).

\subsection{Modelling the effect of HCD absorbers}
\label{sec:modelling}

\begin{figure*}
\includegraphics[width=\textwidth]{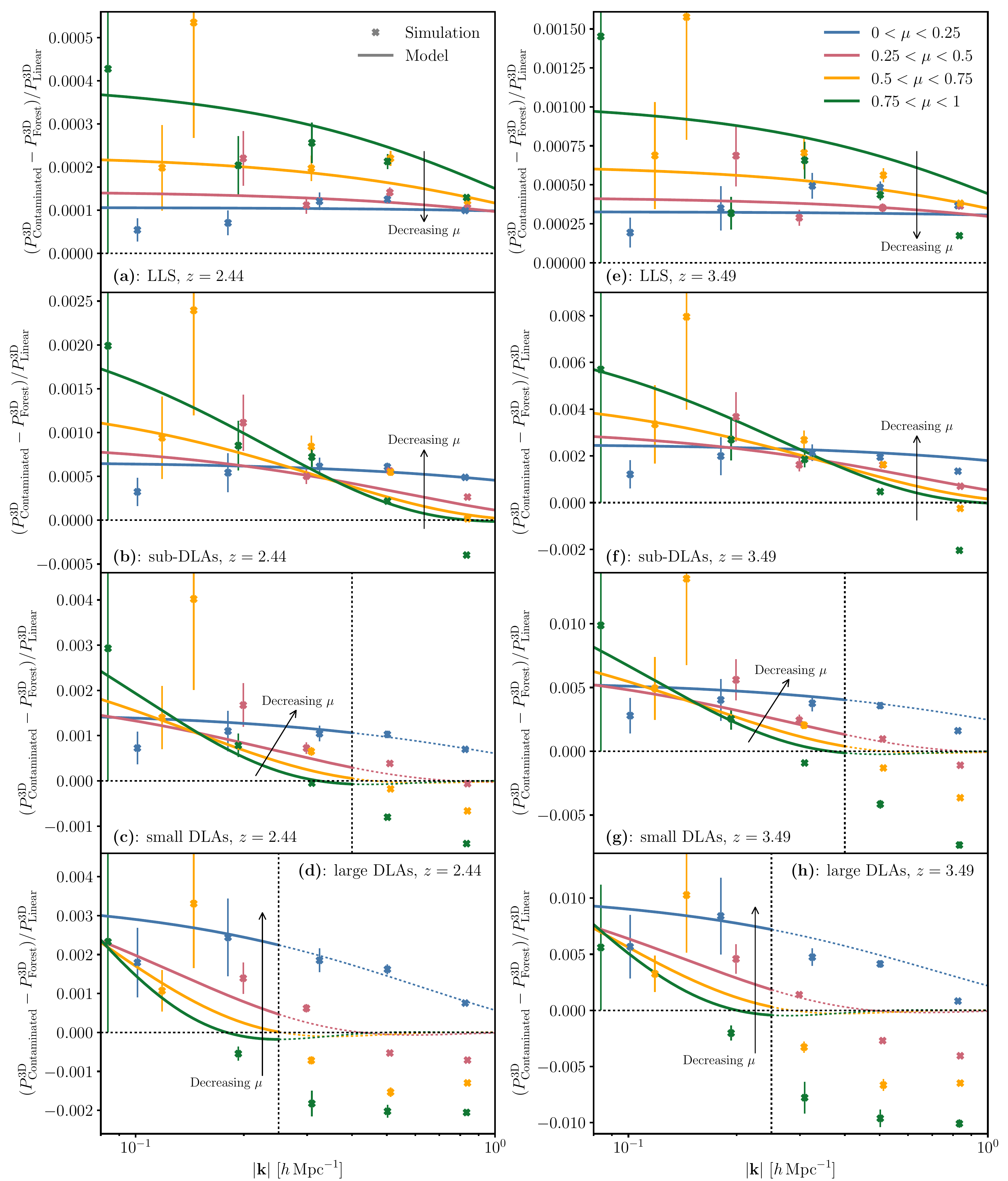}
\caption{The change in the flux power spectrum from contamination of quasar spectra by different categories of HCD absorbers, in ratio to the linear power spectrum. The points are measurements from our simulation boxes; error bars indicate the number of modes in each bin. The lines are maximum posterior values of our preferred model. \textit{From top to bottom}, we show the effect of different categories of HCD absorbers; \textit{from left to right}, we show the effect at different redshifts \(z\). The vertical dotted lines for the two largest HCD absorber categories indicate the smallest scale which we include in our data-vector from our HCD-contaminated simulation boxes for those categories. Our preferred model does not correctly characterise the simulation results for these categories on smaller scales towards the line-of-sight. The dotted lines show an extrapolation of this model, highlighting the discrepancy.}
\label{fig:categories}
\end{figure*}

We show the maximum posterior values of the linear Voigt model and the measurements made in our simulations in Fig.~\ref{fig:categories}. To emphasise the effect of HCD absorbers, we show the part of the model for the auto-correlations of HCD absorbers and their cross-correlation with the Lyman-alpha forest, \ie the last two terms in Eq.~\eqref{eq:power_contam} (\(2 \tilde{b}_\mathrm{Forest} \tilde{b}_\mathrm{HCD} + \tilde{b}^2_\mathrm{HCD}\)). We compare this to the difference between the flux power spectra of the contaminated and uncontaminated boxes of spectra, in ratio to the linear theory matter power spectrum (\((P^\mathrm{3D}_\mathrm{Contaminated} - P^\mathrm{3D}_\mathrm{Forest}) / P^\mathrm{3D}_\mathrm{Linear}\)). We plot the results as a function of column density (by showing the effect for different HCD absorber categories from top to bottom) and as a function of redshift (from left to right). The error bars scale appropriately with the number of modes in each power spectrum bin (\(= d_i \sqrt{2 / N_i}\)), where \(d_i\) is the data-point value, but do not capture the full likelihood function, the details of which are given in Appendix \ref{sec:mcmc_details}.

A measure of the goodness-of-fit is the values of the reduced chi-squared statistic: from top to bottom, left to right, \(\chi^2_\mathrm{red} = \mathrm{(a)}\,1.55\); \(\mathrm{(b)}\,1.56\); \(\mathrm{(c)}\,1.57\); \(\mathrm{(d)}\,1.73\); \(\mathrm{(e)}\,1.04\); \(\mathrm{(f)}\,1.03\); \(\mathrm{(g)}\,1.21\); \(\mathrm{(h)}\,1.24\), all indicating a good fit for the linear Voigt model for all the column densities and redshifts we have considered, excluding certain regimes as explained below. The number of degrees of freedom is (a), (b), (e), (f): 34; (c), (g): 26; (d), (h): 22. The linear Voigt model is discrepant with the simulation results for a small part of the data space (\(|\vec{k}| \gtrsim 0.4\,h\,\mathrm{Mpc}^{-1}\) for small DLAs and \(|\vec{k}| \gtrsim 0.25\,h\,\mathrm{Mpc}^{-1}\) for large DLAs); these exceptions are indicated by the dotted lines in Fig.~\ref{fig:categories} and are restricted to small scales (particularly towards the line-of-sight) for the largest HCD absorbers (small and large DLAs). Indeed, we exclude these parts of the data-vector for the contaminated 3D flux power spectra in our parameter inference; we discuss the implications of this small-scale discrepancy for the highest column density absorbers in \S~\ref{sec:discussion}. Part of the discrepancy between simulation and model in Fig.~\ref{fig:categories} is also driven by the range of \(\mu\) values within each \(\mu\) bin.

For completeness, we quote the maximum (marginalised) posterior values and \(1 \sigma\) credible intervals of the (linear) bias parameters of the Lyman-alpha forest\footnote{Rather than \(b_\mathrm{Forest}\), following \eg \citet{2011JCAP...09..001S,2017arXiv170200176B}, we sample the combination \(b_\mathrm{Forest} (1 + \beta_\mathrm{Forest})\), which is less correlated with \(\beta_\mathrm{Forest}\).} at \(z = 2.44\):
\begin{equation*}
b_\mathrm{Forest} (1 + \beta_\mathrm{Forest}) = -0.270 \pm 0.004;\,\,\,\,\, \beta_\mathrm{Forest} = 1.722 \pm 0.072
\end{equation*}
and at \(z = 3.49\):
\begin{equation*}
b_\mathrm{Forest} (1 + \beta_\mathrm{Forest}) = -0.511 \pm 0.006;\,\,\,\,\, \beta_\mathrm{Forest} = 1.249 \pm 0.043.
\end{equation*}
The posterior on \(\beta_\mathrm{Forest}\) at \(z = 2.44\) is in \(1 \sigma\) agreement with the best-fit value from BOSS DR12 spectra \citep{2017arXiv170200176B}, \(\beta^\mathrm{BOSS}_\mathrm{Forest} = 1.663 \pm 0.085\) at a central redshift of \(z = 2.3\). However, the posterior on \(b_\mathrm{Forest} (1 + \beta_\mathrm{Forest})\) is lower than the value measured from data \(b^\mathrm{BOSS}_\mathrm{Forest} (1 + \beta^\mathrm{BOSS}_\mathrm{Forest}) = -0.325 \pm 0.004\) at \(z = 2.3\); this difference has been observed in other studies with hydrodynamical simulations \citep[\eg][]{2015JCAP...12..017A}. The redshift evolution in \(b_\mathrm{Forest}\) observed in our simulations (modelled as \(b_\mathrm{Forest} \propto (1 + z)^\gamma\)) implies \(\gamma = 3.1\), roughly matching the value currently assumed in data analyses \(\gamma^\mathrm{BOSS} = 2.9\) (see \S~\ref{sec:forest_theory}). We find that \(\beta_\mathrm{Forest}\) decreases at higher redshift, also as observed in previous studies with simulations \citep{2015JCAP...12..017A}. In Appendix \ref{sec:test_bias}, we test the sensitivity of our inference of the bias parameters of the Lyman-alpha forest to the smallest scale included in our analysis \(|\vec{k}|_\mathrm{max}\); we find that our inferences are overall insensitive to this, suggesting that our results are robust to our modelling of non-linear effects. We also recover the same posterior distributions on the Lyman-alpha forest bias parameters for each of the HCD absorber categories of contaminated flux power spectra that we consider. These parameters also match those inferred from the 3D Lyman-alpha forest flux power spectrum only.

As discussed in \S~\ref{sec:mcmc_modelling}, we place Gaussian priors on the bias parameters of the different categories of HCD absorber (but not the forest biases), which are otherwise poorly constrained, since the amplitude of their effect is sub-dominant to the Lyman-alpha forest flux power spectrum. These prior distributions are returned almost exactly in the marginalised posteriors. The scale-dependence of the effect of HCD absorbers, meanwhile, is fully determined by the physics of their absorption profiles and the appropriate CDDF.

\begin{figure}
\includegraphics[width=\columnwidth]{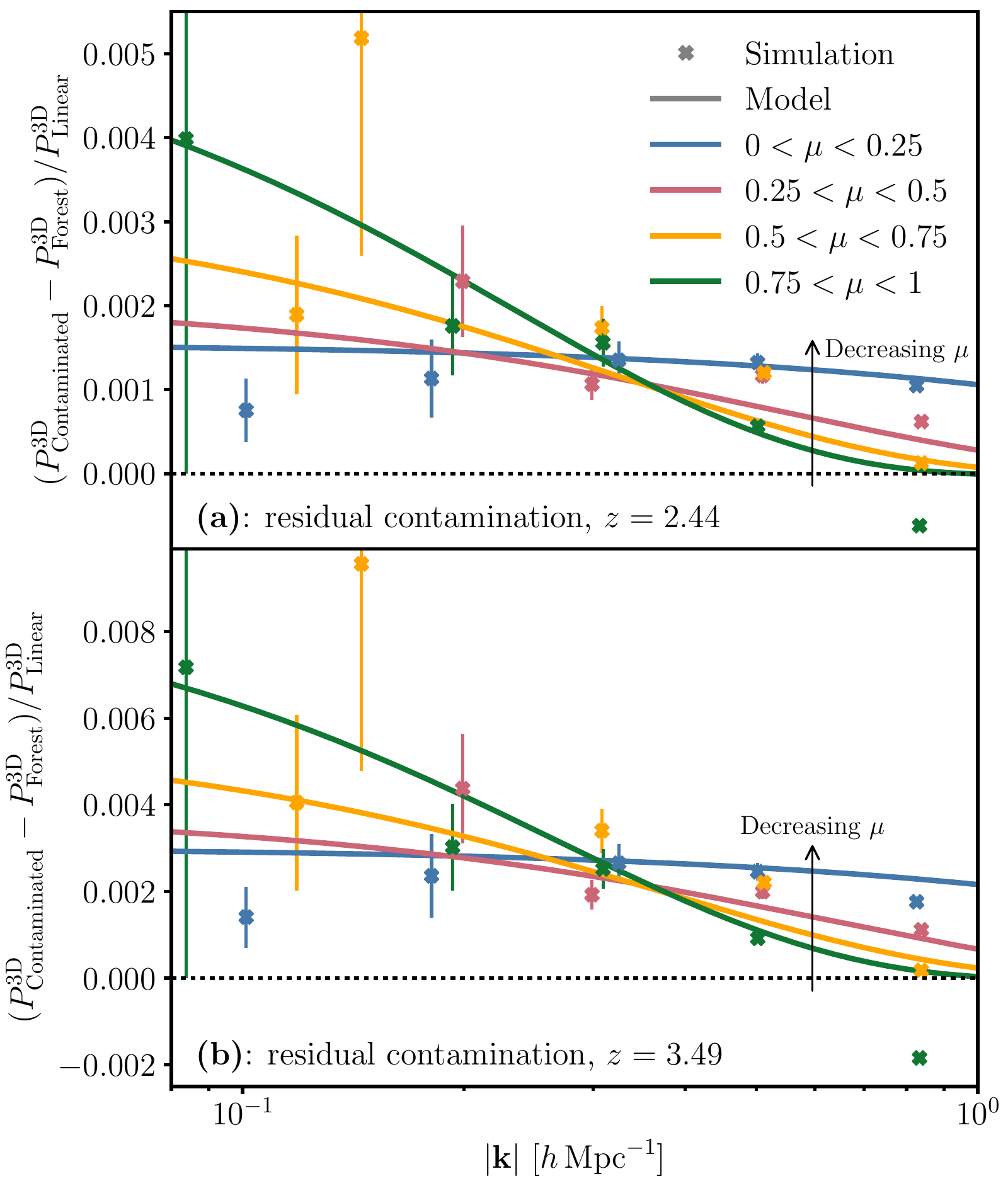}
\caption{As Fig.~\ref{fig:categories}, but for a mock residual contamination after the largest HCD absorbers have been ``clipped'' from quasar spectra (\ie only LLS and sub-DLAs remaining). \textit{From top to bottom}, we show the effect at different redshifts \(z\).}
\label{fig:residual}
\end{figure}

Figure \ref{fig:residual} compares the maximum posterior values of the linear Voigt model to the effect of a mock residual contamination of HCD absorbers on the 3D flux power spectrum, at the two redshifts we consider. This mock residual contamination (of LLS and sub-DLAs) approximately matches the column densities assumed remaining in BOSS spectra \citep{2017arXiv170200176B} after the largest damping wings have been removed. We find that the linear Voigt model is in statistical agreement with our simulation measurements (\(\chi^2_\mathrm{red} =\,\mathrm{(a)}\,1.56\); \(\mathrm{(b)}\,1.03\)). The number of degrees of freedom is (for both panels) 34. We highlight this configuration because, although Fig.~\ref{fig:categories} shows that there are some small scales towards the line-of-sight for the largest HCD absorbers where the simple linear Voigt model is not appropriate, these are the HCD absorber categories most efficiently removed in the ``clipping'' process in data analysis.

\section{Discussion}
\label{sec:discussion}

In \S~\ref{sec:modelling}, we showed the regimes in scale and column density where the simple linear Voigt model (see \S~\ref{sec:hcd_theory}) can characterise the effect of HCD absorber contamination on correlations in the 3D Lyman-alpha forest. The linear Voigt model is arguably the simplest model that can be constructed to take account of the true absorption line profiles of HCD absorbers. It is a linear model with HCD absorbers as biased tracers of the matter density distribution with redshift-space distortions. This is then convolved with the Voigt profiles of HCD absorbers' damping wings and integrated over the CDDF of the absorbers. The linear Voigt model manifests as a suppression in power due to damping wings which remove structure in the spectra. This effect is stronger towards the line-of-sight since this is the direction in which the wings appear. The scales at which the suppression starts (and the overall amplitude) are larger for more dense absorber categories since their wings are wider. The suppression is such that it overcomes the boost in power towards the line-of-sight on large scales due to redshift-space distortions; consequently, there is a characteristic cross-over in the different curves for each absorber category. The effect of HCD absorbers transverse to the line-of-sight is scale-independent, since there is no component of the damping wings in this direction (our lowest \(\mu\) bin does include some modes slightly away from the transverse direction). The amplitude of the effect increases with redshift, mainly because the cross-correlation with the Lyman-alpha forest is stronger (there being overall more absorption at higher redshift).

However, on small scales towards the line-of-sight for the largest HCD absorber categories (\(|\vec{k}| \gtrsim 0.4\,h\,\mathrm{Mpc}^{-1}\) for small DLAs and  \(|\vec{k}| \gtrsim 0.25\,h\,\mathrm{Mpc}^{-1}\) for large DLAs), the linear Voigt model cannot characterise our simulation results. Power is suppressed towards the line-of-sight more strongly than our model allows such that there is less power than without the HCD absorbers. We consider two possible causes of this discrepancy with the linear Voigt model (as mentioned in \S~\ref{sec:hcd_theory}). First, our model does not consider any non-linear clustering of the gas or halos associated with HCD absorbers. A comprehensive model for the clustering of HCD absorbers should certainly account for this effect. However, we found no preference for a parametric form \citep[akin to that used for the Lyman-alpha forest;][]{2003ApJ...585...34M,2015JCAP...12..017A} that would improve the fit to our simulation results. This suggests that such a closed form cannot alone account for the discrepancy. Second, we consider the non-linear effect of a three-point correlation between a Lyman-alpha forest fluctuation and an HCD absorber fluctuation at the same position and a Lyman-alpha forest fluctuation at a second position (see \S~\ref{sec:hcd_theory}). This term was shown by \citet{2012JCAP...07..028F} to be at least as significant as the HCD absorber auto-correlation on small scales (separations \(r < 40\,\mathrm{Mpc}\,h^{-1}\)) and to have the correct (negative) sign to account for the additional suppression of power observed on small scales towards the line-of-sight for small and large DLAs. (This is because it is the correlation between three negatively-biased tracers.) It is intuitively understood as the effect of the damping wings in masking regions of the Lyman-alpha forest and so suppressing auto-correlations in the Lyman-alpha forest that would otherwise occur on scales within the widths of individual wings. This effect will be stronger for more dense HCD absorbers since their damping wings are wider and so mask more of the Lyman-alpha forest; stronger on scales smaller than the widths of wings; and stronger towards the line-of-sight since this is the direction in which the masking occurs. This seems a qualitative match to the observed discrepancies with the linear Voigt model, but as yet there exists no simple model for this higher-order effect and we have not explicitly tested whether it can account for the observed discrepancies. As discussed above and in \S~\ref{sec:modelling}, the effect is restricted to the highest column densities, which are in any case mostly removed in the clipping pre-processing of spectra.

\begin{figure}
\includegraphics[width=\columnwidth]{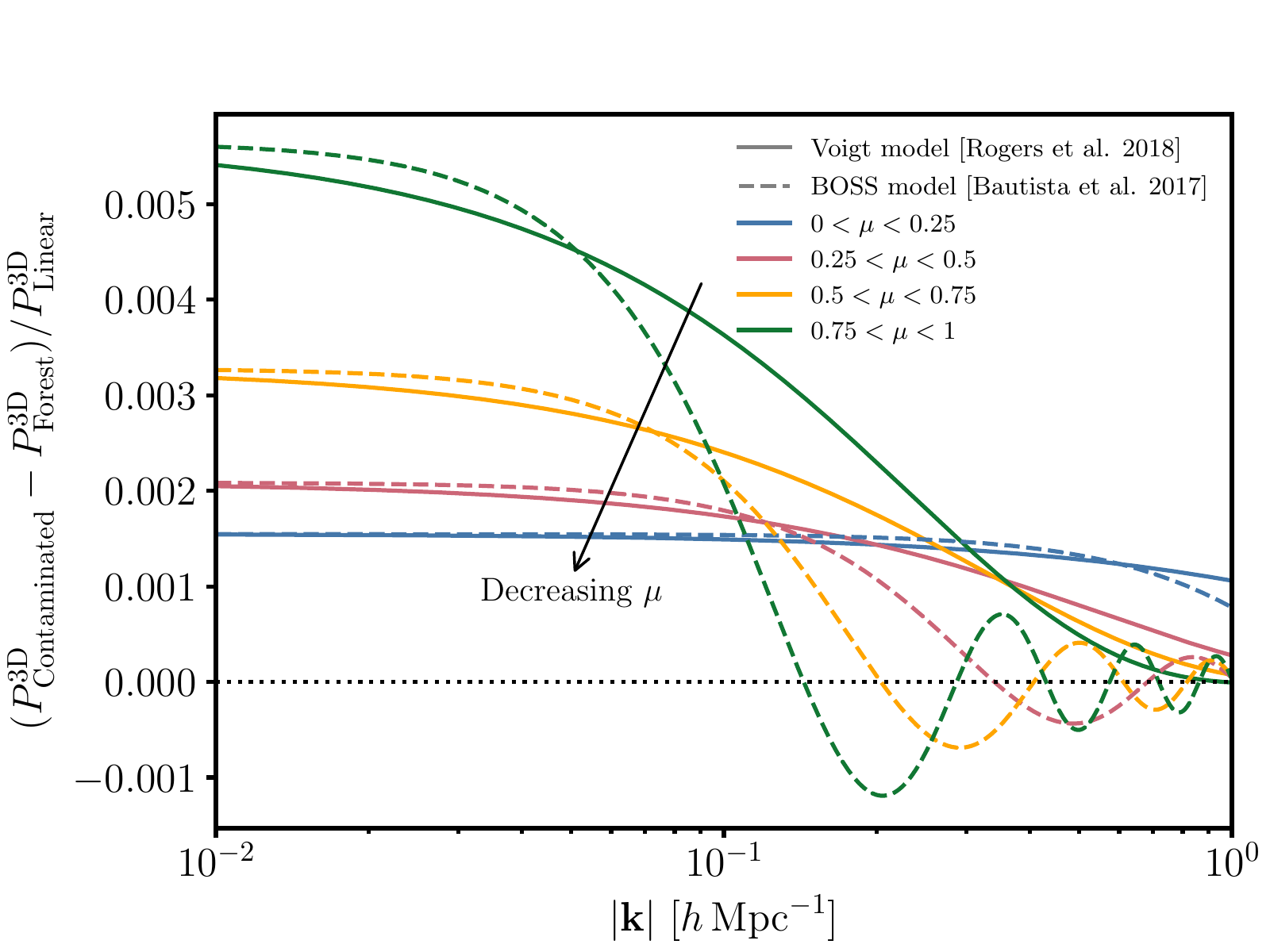}
\caption{A comparison of the existing model as used by the BOSS Collaboration \citep{2017arXiv170200176B,2017arXiv170900889P} and the linear Voigt model presented in this study for the additive effect of residual HCD absorbers (after the ``clipping'' of the largest absorbers from quasar spectra) on the three-dimensional flux power spectrum. For the linear Voigt model, we show the maximum posterior values as inferred from a mock residual contamination in our simulation box at \(z = 2.44\). For the BOSS model, we rescale to match the bias and redshift-space distortions inferred in our box, but use the best-fit value of the shape parameter as found in BOSS mock spectra with a residual contamination and data. The maximum posterior value of the BOSS model as inferred from our simulation gives unphysical results on scales larger than the size of our box (see Fig.~\ref{fig:comparison_sim_fits}).}
\label{fig:comparison}
\end{figure}

\begin{figure}
\includegraphics[width=\columnwidth]{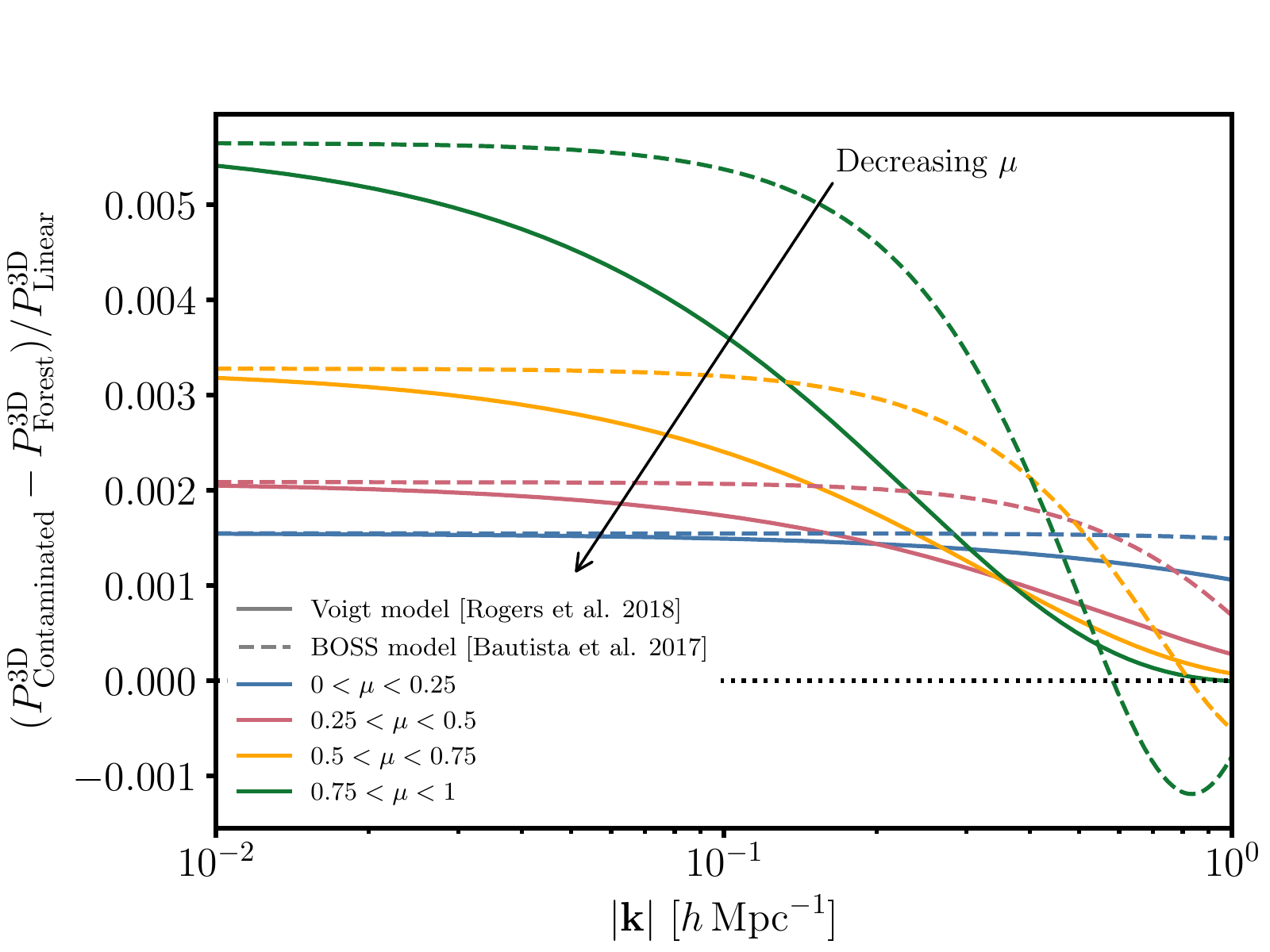}
\caption{As Fig.~\ref{fig:comparison}, but the BOSS model is shown for the maximum posterior value of the shape parameter as inferred from our simulation. This highlights how the BOSS model, in attempting to fit the small-scale behaviour in our box, gives a flat response on scales larger than the size of our box. This flat response suggests there is no large-scale effect of HCD absorbers. This contradicts the physical linear Voigt model and the results from data and mock spectra from the BOSS Collaboration \citep{2017arXiv170200176B,2017arXiv170802225D,2017arXiv170900889P}.}
\label{fig:comparison_sim_fits}
\end{figure}

Figure \ref{fig:comparison} compares the linear Voigt model as inferred from our simulations contaminated by a mock residual contamination of HCD absorbers (only LLS and sub-DLAs remaining, approximating the effect of clipping out the damping wings of more dense absorbers as is done with survey spectra), with the model used by the BOSS Collaboration \citep{2017arXiv170200176B,2017arXiv170802225D} for the same effect. This ``BOSS model'' approximates the damping wings as top-hats and so the effect on the Fourier space correlations (\ie the flux power spectrum) is a sinc function (see \S~\ref{sec:hcd_theory}). We rescale the BOSS model to have the same bias and redshift-space distortions as inferred in our simulation box (for a fair comparison to the linear Voigt model), but use the shape parameter as found in BOSS mock spectra with a residual contamination and in data (\(L_\mathrm{HCD} = 24.341\,\mathrm{Mpc}\,h^{-1}\))\footnote{We use this shape parameter value because the shape parameter of the BOSS model as inferred from our simulation box was considerably smaller than the BOSS best-fit value (in order to fit the small-scale correlations) and gave a flat response on scales larger than the size of our box (\ie indicating no effect of damping wings in contradiction to the physical linear Voigt model; see Fig.~\ref{fig:comparison_sim_fits}).}. We extrapolate the BOSS model to smaller scales than considered in their analysis, where the minimum separations measured were \(r = 10\,\mathrm{Mpc}\,h^{-1}\). Although our inference on the linear Voigt model is only constrained by the scales accessible in our simulation box, we extrapolate this model to larger scales of relevance to a BAO analysis. Although we are not able to explicitly test the model on these larger scales, it is expected to correctly characterise the effect as it constitutes the physical expectation on large scales. We conclude from Fig.~\ref{fig:comparison} that the BOSS model constitutes a good approximation for scales of relevance for a BAO analysis, but that on smaller scales, the linear Voigt model should be used in order to account for the effect of extended damping wings for a residual contamination of HCD absorbers. There is always a residual contamination because it is difficult to identify narrow damping wings amongst the superposed Lyman-alpha forest absorption lines and instrumental noise. This becomes harder for noisier quasar spectra. By definition, it is impossible to know the exact efficiency of detection algorithms on survey data, although this can be estimated on simulations. However, the model we have constructed (as given by Eq.~\eqref{eq:power_extra}) has the flexibility to marginalise over this uncertainty, by sub-dividing HCD absorbers into categories, each with their own set of parameters. Appropriate priors reflecting the expected down-weighting of higher column density systems can be constructed based on previous analyses and simulation testing; the details will be survey-specific.


\section{Conclusions}
\label{sec:concs}

We have measured the effect of contamination of quasar spectra by the damping wings of high column density (HCD) absorbing regions of neutral hydrogen on correlations in the 3D Lyman-alpha forest. We accomplished this by measuring 3D flux power spectra from a cosmological hydrodynamical simulation \citep[Illustris;][]{2014Natur.509..177V,2015A&C....13...12N} as a function of the column density of the HCD absorber contamination and redshift. We found that, even after the largest damping wings have been removed (as performed by survey pipelines), that the effect of the residual contamination can be as large as a 15\% correction to the 3D Lyman-alpha forest flux power spectrum (at \(|\vec{k}| = 0.1\,h\,\mathrm{Mpc}^{-1}\)). We found that the effect of this residual contamination can be characterised by a simple linear model (with bias and redshift-space distortions) convolved with the Voigt profiles of the damping wings and integrated over the column density distribution function of the HCD absorbers. This model also successfully characterises the contamination effect on large scales for the highest column densities; however, on smaller scales (\eg \(|\vec{k}| > 0.4\,h\,\mathrm{Mpc}^{-1}\) for small DLAs) towards the line-of-sight, the model fails possibly due to additional suppression in power by the most massive systems due to the effective masking of auto-correlations in the Lyman-alpha forest by their damping wings. \citet{FontRiberaP3D} found that there is much more constraining power in the 3D flux power spectrum than the 1D power spectrum for BOSS for \(|\vec{k}| < 1\,h\,\mathrm{Mpc}^{-1}\), underlying the importance of accurately modelling systematics up to small scales. We therefore find that this linear Voigt model will help with precision measurements of BAO in future surveys (eBOSS/DESI) and will be essential for reconstructing the power spectrum shape beyond BAO.

\section*{Acknowledgements}
\label{sec:ack}

KKR, SB, HVP and BL thank the organisers of the COSMO21 symposium in 2016, where this project was devised. KKR was supported by the Science and Technology Facilities Council (STFC). SB was supported by NASA through Einstein Postdoctoral Fellowship Award Number PF5-160133. HVP was partially supported by the European Research Council (ERC) under the European Community's Seventh Framework Programme (FP7/2007-2013)/ERC grant agreement number 306478-CosmicDawn. AP was supported by the Royal Society. AFR was supported by an STFC Ernest Rutherford Fellowship, grant reference ST/N003853/1. BL was supported by NASA through Einstein Postdoctoral Fellowship Award Number PF6-170154. This work was partially enabled by funding from the University College London (UCL) Cosmoparticle Initiative.

\bibliographystyle{mymnras_eprint}
\bibliography{dla_bias_3D}

\appendix
\section{Tests of HCD absorber dodging}
\label{sec:test_dodging}

In this Appendix, we test the effect of replacing simulated spectra contaminated by HCD absorption with nearby uncontaminated spectra on our measurements of 3D flux power spectra. The measurements of these power spectra are made computationally simple by the use of FFTs, which in turn require a regular grid of samples. However, the transverse HCD absorber ``dodging'' of some spectra makes this grid irregular. An error therefore arises from treating this irregular grid as the original regular grid (\ie to ignore the transverse dodging distances) in computing the necessary FFTs.

\begin{figure}
\includegraphics[width=\columnwidth]{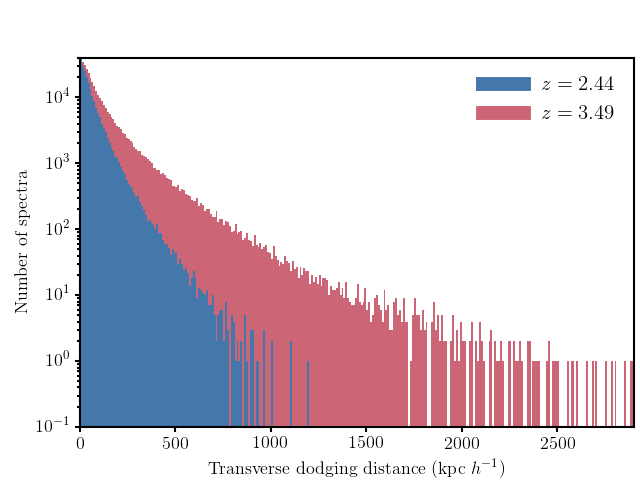}
\caption{Histogram of the transverse comoving distances ``dodged'' by each simulated spectrum in order to avoid HCD absorbers. The total number of spectra at each redshift \(z\) is 562,500. The number of spectra remaining un-dodged at \(z = 2.44\) and \(3.49\) is respectively 391,500 (69.6\%) and 257,063 (45.7\%). There is a tail of large dodging distances, much larger than the physical size of the most massive HCD absorbers because sometimes, in dodging one absorber, the proposed replacement spectrum will coincide with another absorber, somewhere else along the line-of-sight, requiring further dodging.}
\label{fig:dodging_statistics}
\end{figure}

Figure \ref{fig:dodging_statistics} shows the distribution of the transverse dodging distances required to find replacement mock spectra uncontaminated by HCD absorbers, for the simulation boxes at the two redshifts we consider. (See \S~\ref{sec:hcd_dodging} for more details about why and how we dodge HCD absorbers.) Replacement spectra are trialled increasingly further away from the original spectrum in steps of \(10\,\mathrm{kpc}\,h^{-1}\) until an uncontaminated spectrum is found. Many of the final replacement spectra require many iterations to be found; this is exacerbated by dodging one HCD absorber but then finding another HCD absorber elsewhere along the spectrum which then requires further dodging. More dodging is required at higher redshift because neutral hydrogen is more abundant.

\begin{figure}
\includegraphics[width=\columnwidth]{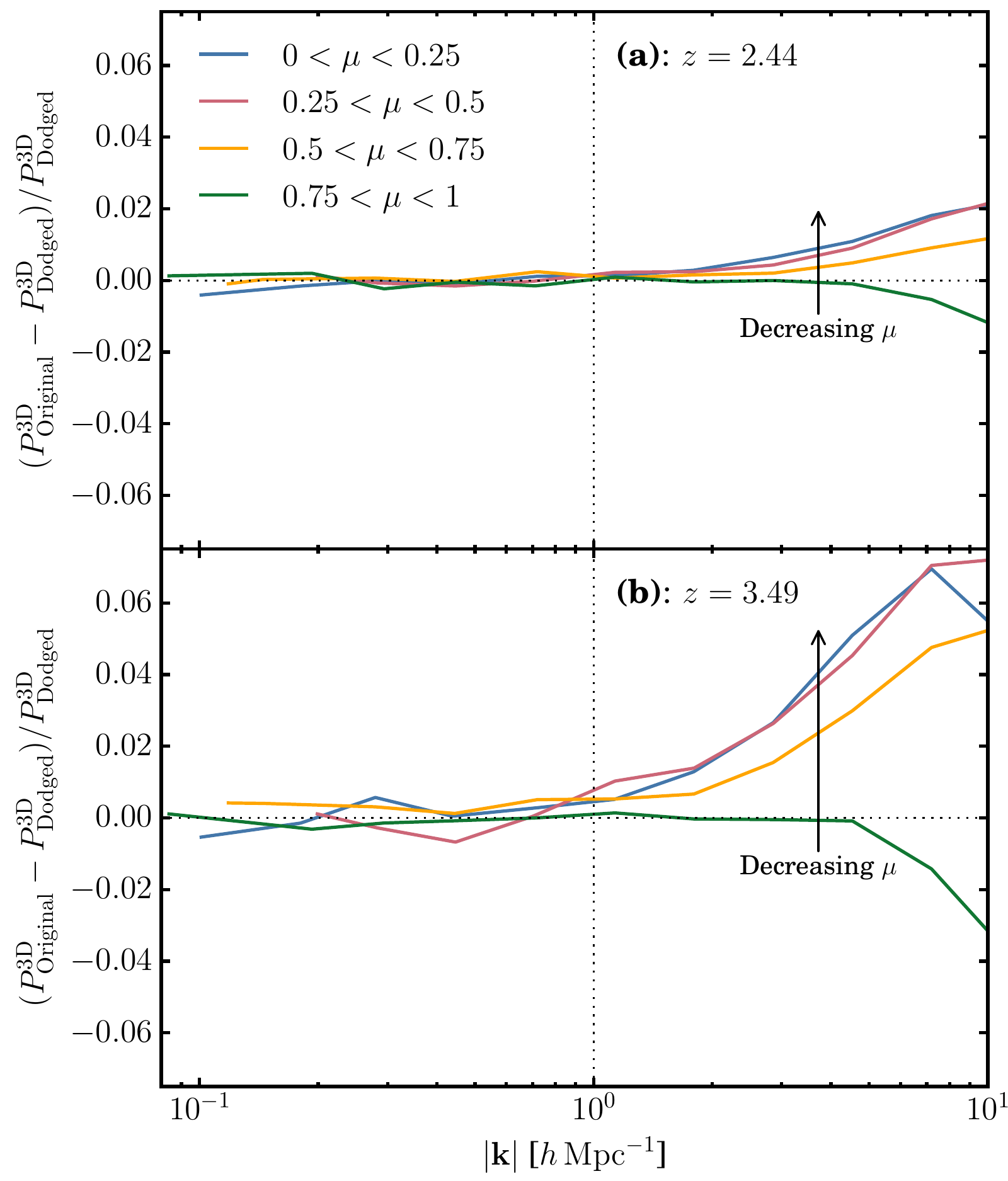}
\caption{The fractional error in the estimation of the power spectrum of a Gaussian random field (GRF) due to ``dodging'' lines-of-sight. We replicate in a GRF the exact movement of lines-of-sight that we carry out in our simulation box in order to dodge HCD absorbers. We then calculate the error in the estimation of the power spectrum due to ignoring the changes in position of lines-of-sight when calculating the necessary (fast) Fourier transforms. We note that the error remains small (sub-percent) for scales of interest in our study (\(|\mathbf{k}| < 1\,h\,\mathrm{Mpc}^{-1}\)). \textbf{(a)} \textit{above}: we replicate the dodging in our box at \(z = 2.44\); \textbf{(b)} \textit{below}: at \(z = 3.49\).}
\label{fig:fractional_dodging_effect}
\end{figure}

We have tested the effect of moving some lines-of-sight, but then ignoring the changes in positions in the calculation of FFTs. In order to approximate our box of mock spectra, we generate a Gaussian random field (GRF) from a cosmological power spectrum (the same as input to the simulations). It is sampled at the same resolution as our mock spectra. We then estimate the power spectrum from this box, in the same way that we do with our mock spectra, forming \(P_\mathrm{Original}^\mathrm{3D}\). In order to approximate our box of mock spectra after the dodging procedure, we then replicate on our GRF the exact movement of lines-of-sight as we carry out in our simulations. We then estimate the power spectrum from this new ``dodged'' box, in the same way that we do with our mock spectra, \ie ignoring the changes in positions of our replacement lines-of-sight. This forms \(P_\mathrm{Dodged}^\mathrm{3D}\). An error is introduced in estimating the power spectrum and ignoring the transverse distances that some of the samples of the field have moved (this error could be avoided if we didn't rely in the calculation of power spectra on FFTs which require evenly sampled functions). Figure \ref{fig:fractional_dodging_effect} shows the fractional error from this effect for the dodges we carry out at the different redshifts we consider. There is more error at higher redshift because there is more dodging (see Fig.~\ref{fig:dodging_statistics}), but the error remains sub-percent for scales of interest for our study (\(|\mathbf{k}| < 1\,h\,\mathrm{Mpc}^{-1}\)). This figure should be compared to Fig.~\ref{fig:fractional_hcd_effect} \textit{(b)} (the fractional effect of a mock residual contamination of HCD absorbers on the 3D Lyman-alpha forest flux power spectrum); for the scales of interest, the effect of the HCD absorbers that we wish to measure remains much larger than the error arising from the dodging. We therefore ignore only scales smaller than this cut-off in our analysis.

\section{Details of modelling and MCMC sampling}
\label{sec:mcmc_details}

In \S~\ref{sec:mcmc_modelling}, we wish to sample the joint posterior probability distribution of the parameters of our proposed model in Eq.~\eqref{eq:power_contam}, given our measured 3D flux power spectra. We therefore require a likelihood function for our simulated data given the model. We use a Gaussian likelihood function and assume the covariance matrix to be diagonal (\ie we ignore correlations between power spectrum bins). Each \(\hat{\delta}_\mathrm{Flux}(\vec{k}_n)\) is well approximated by a Gaussian random variable and so, as explained in \S~\ref{sec:pow_spectrum}, our flux power spectrum estimates in each bin \(i\) are the sums of the squares of \(N_i\) such variables. It follows that each of the elements of our data-vector is chi-squared distributed (ignoring the slightly different amplitudes of and possible correlations between the Fourier modes within each bin):\footnote{For practical purposes, for \(N_i > 50\), the distribution is close to a Gaussian distribution.}
\begin{equation}\label{eq:data_distribution}
P^\mathrm{3D}_{\mathrm{Flux},i} \sim \frac{P^\mathrm{3D}_{\mathrm{True},i}}{N_i} \chi^2(N_i),
\end{equation}
where \(P^\mathrm{3D}_{\mathrm{True},i}\) is the true (ensemble) value of the 3D flux power spectrum in bin \(i\). The variance of this distribution is \(2 (P^\mathrm{3D}_{\mathrm{True},i})^2 / N_i\) and these form the (diagonal) elements of our covariance matrix (substituting \(P^\mathrm{3D}_{\mathrm{Flux},i}\) for \(P^\mathrm{3D}_{\mathrm{True},i}\), which is otherwise \textit{a priori} unknown). To constrain the contamination parameters while marginalising over intrinsic Lyman-alpha forest bias parameters, we combine the likelihoods for the \(P^\mathrm{3D}_\mathrm{Contaminated}\) and \(P^\mathrm{3D}_\mathrm{Forest}\) data-vectors (using Eq.~\eqref{eq:forest_linear} to model \(P^\mathrm{3D}_\mathrm{Forest}\)); we ignore correlations between the two data-vectors and simply add the log-likelihoods. This is sufficient for the level of accuracy of our study.

Our model requires evaluation of \(P^\mathrm{3D}_\mathrm{Linear}\) in each bin; to improve the comparison to \(P^\mathrm{3D}_{\mathrm{Flux},i}\), we similarly evaluate \(P^\mathrm{3D}_\mathrm{Linear}\) at each individual mode and bin in the same way. We associate with each bin the average values of \(|\vec{k}|\) and \(\mu\) from the contributing modes. We use uniform prior probability distributions for the Lyman-alpha forest bias parameters \(b_\mathrm{Forest}\) and \(\beta_\mathrm{Forest}\) and use Gaussian priors for the HCD absorber bias parameters \(b_\mathrm{HCD}\) and \(\beta_\mathrm{HCD}\), which are otherwise poorly constrained. The mean and \(1 \sigma\) values of the Gaussian priors on \(b_\mathrm{HCD}\) are (following the best-fit values on the total \(b_\mathrm{HCD}\) from \citealt{2017arXiv170200176B}) \textit{from top to bottom, left to right} in Fig.~\ref{fig:categories}, \(\mathrm{(a)}\,-0.0005 \pm 0.0002\); \(\mathrm{(b)}\,-0.003 \pm 0.001\); \(\mathrm{(c)}\,-0.007 \pm 0.003\); \(\mathrm{(d)}\,-0.016 \pm 0.006\); \(\mathrm{(e)}\,-0.0007 \pm 0.0002\); \(\mathrm{(f)}\,-0.005 \pm 0.002\); \(\mathrm{(g)}\,-0.012 \pm 0.004\); \(\mathrm{(h)}\,-0.022 \pm 0.009\). They are estimated by dividing the total \(b_\mathrm{HCD}\) from \citet{2017arXiv170200176B} by the relative rest-frame equivalent widths of the damping wings of each absorber category; they are scaled up at higher redshift by the increased amount of HCD absorption (estimated from the fraction of contaminated spectra). Following the best-fit values found by \citet{2017arXiv170200176B}, we place a Gaussian prior on \(\beta_\mathrm{HCD} = 0.7 \pm 0.2\). Our prior distributions for these contamination parameters are almost exactly returned in their marginalised 1D posterior distributions (details are given in \S~\ref{sec:modelling}), \ie we are very insensitive to the amplitude of the effect of HCD absorbers. The shape of the scale-dependent bias is fully determined by the physics of the Voigt absorption profiles and the CDDF (see \S~\ref{sec:hcd_theory}). We investigate the suitability of our model for the scale-dependent bias arising from the absorption profiles of HCD absorbers as a function of column density by repeating the posterior sampling for the \(P^\mathrm{3D}_\mathrm{Contaminated}\) constructed for each HCD absorber category.

We sample the posterior distributions using a Markov chain Monte Carlo method, specifically \texttt{emcee} \citep{2013PASP..125..306F}, an implementation of the affine-invariant MCMC sampler. We initialise our chains uniformly within the (non-zero) bounds of our prior distributions and test for convergence using the Gelman-Rubin statistic \citep{gelman1992,doi:10.1080/10618600.1998.10474787}.

\section{Tests of robustness of inference of bias parameters of the Lyman-alpha forest}
\label{sec:test_bias}

\begin{figure}
\includegraphics[width=\columnwidth]{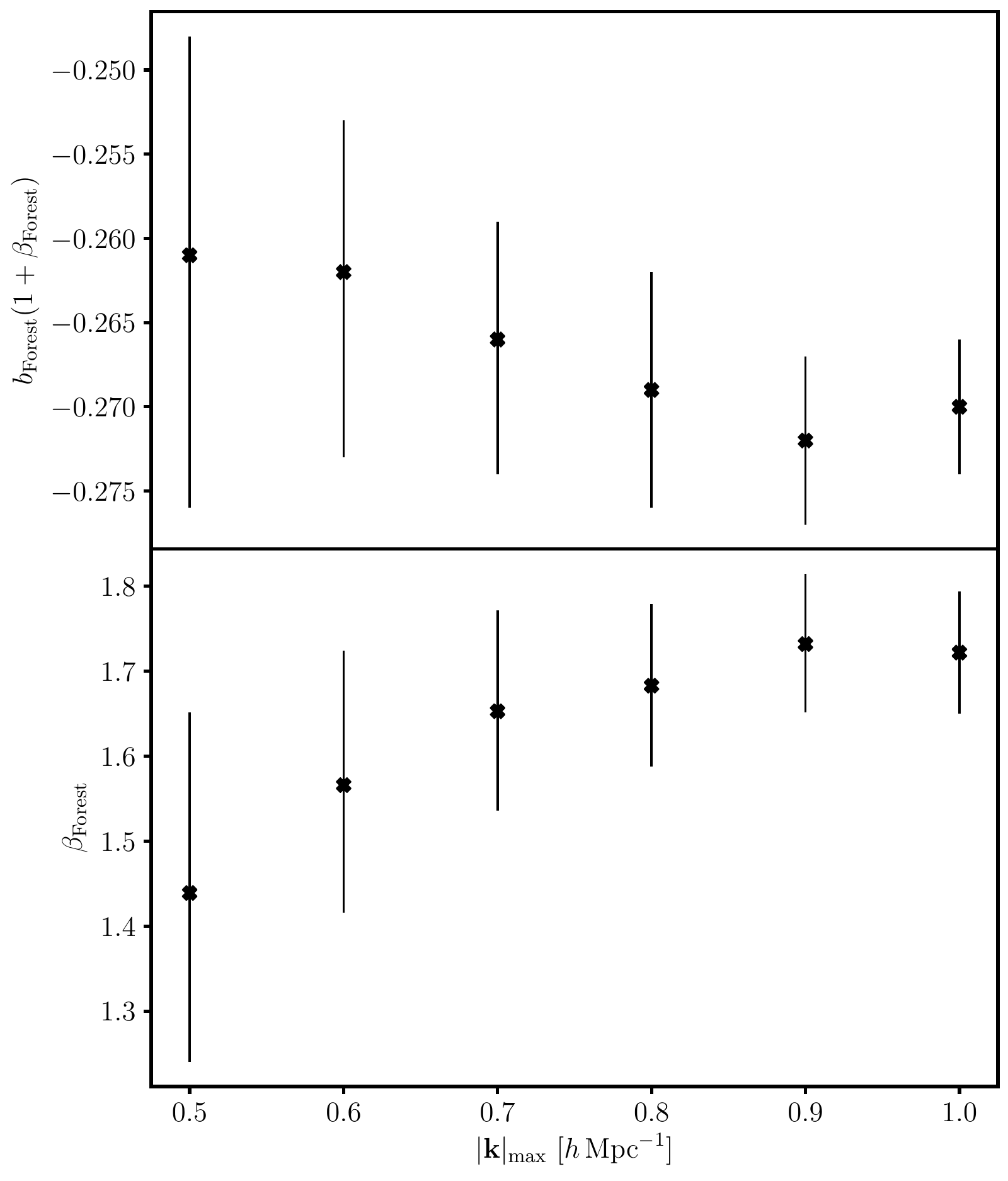}
\caption{The maximum posterior values with the \(1 \sigma\) credible intervals of the bias \(b_\mathrm{Forest}\) (\textit{above}) and redshift space distortion \(\beta_\mathrm{Forest}\) (\textit{below}) parameters of the Lyman-alpha forest, as inferred from our simulation box at \(z = 2.44\), using different values of \(|\vec{k}|_\mathrm{max}\), the smallest scale included in our analysis. Following \eg \citet{2011JCAP...09..001S,2017arXiv170200176B}, we sample the combination \(b_\mathrm{Forest} (1 + \beta_\mathrm{Forest})\), which is less correlated with \(\beta_\mathrm{Forest}\). We find that our marginalised parameter posteriors are statistically consistent, irregardless of the smallest scale at which we cut our data vector.}
\label{fig:bias_tests}
\end{figure}

Figure \ref{fig:bias_tests} shows the results of testing how changing the smallest scale that we include in our analysis \(|\vec{k}|_\mathrm{max}\) affects the (marginalised 1D) posterior distributions inferred for the bias parameters of the Lyman-alpha forest. The \(1 \sigma\) credible intervals on the combination \(b_\mathrm{Forest} (1 + \beta_\mathrm{Forest})\) and \(\beta_\mathrm{Forest}\) increase as \(|\vec{k}|_\mathrm{max}\) decreases because the number of modes remaining on scales larger than \(|\vec{k}|_\mathrm{max}\) falls off quite sharply as \(|\vec{k}|_\mathrm{max}\) is reduced. Although the combination \(b_\mathrm{Forest} (1 + \beta_\mathrm{Forest})\) is sampled, rather than \(b_\mathrm{Forest}\) alone, because it is less correlated with \(\beta_\mathrm{Forest}\), there is evidently still correlation: as \(\beta_\mathrm{Forest}\) decreases with \(|\vec{k}|_\mathrm{max}\), so does also the amplitude of \(b_\mathrm{Forest} (1 + \beta_\mathrm{Forest})\). Nonetheless, the posteriors of both bias parameters are statistically consistent for all the values of \(|\vec{k}|_\mathrm{max}\) that we consider, suggesting a degree of robustness in our inference on these parameters. Moreover, our conclusions on the scale-dependence of the effect of HCD absorbers on correlations in the 3D Lyman-alpha forest are insensitive to the overall amplitude, including the biases of the Lyman-alpha forest.

\end{document}